\documentclass[onecolumn, a4paper, times, trackchanges]{aastex63}
\usepackage[english]{babel}

\usepackage{amsmath} 

\usepackage{graphicx}
\usepackage{natbib}

\newcommand{\tot}{_{\text{tot}}}
\newcommand{\E}{_{\text{E}}}

\newcommand{\lya}{Lyman-$\alpha$~{}}

\newcommand{\myvec}[1]{\ensuremath{\boldsymbol{{#1}}}}

\begin{document}
\title{Absorption of the \lya radiation in the heliosphere}

\correspondingauthor{I. Kowalska-Leszczynska}
\email{ikowalska@cbk.waw.pl}

\author[0000-0002-6569-3800]{I. Kowalska-Leszczynska}
\affil{Space Research Centre PAS (CBK PAN),\\
Bartycka 18A, 00-716 Warsaw, Poland} 

\author[0000-0002-5204-9645]{M. A. Kubiak}
\affil{Space Research Centre PAS (CBK PAN),\\
Bartycka 18A, 00-716 Warsaw, Poland}

\author[0000-0003-3957-2359]{M. Bzowski}
\affil{Space Research Centre PAS (CBK PAN),\\
Bartycka 18A, 00-716 Warsaw, Poland}

\begin{abstract}
Absorption of the \lya{} radiation on Interstellar Neutral Hydrogen (ISN H) atoms in the heliosphere is a potentially important effect to account for in precise gas distribution simulations.
In this paper we develop a method to estimate the magnitude of absorption of solar \lya radiation inside the solar wind termination shock and to include absorption effects in the Warsaw Test Particle Model (WTPM) by an appropriate modification of radiation pressure.
We perform calculations of absorption effects on a 3D grid in the heliosphere and present a set of parameters to model absorption effects for the mean solar activity conditions.
We show that absorption can change up to 3\% depending on the solar activity level.
Using a modified version of WTPM, we calculate the expected signal from IBEX-Lo and show that absorption may modify the simulated flux up to 8\%. 
\end{abstract}

\section{Introduction}
\label{sec:intro}
\noindent

The heliosphere is created due to interaction between the solar wind and the magnetized, partly ionized interstellar gas surrounding the Sun.
The neutral component, composed mostly of neutral hydrogen and further on referred to as interstellar neutral hydrogen (ISN H), penetrates freely inside the heliosphere.
Since the Sun is moving relative to the local interstellar matter, ISN H creates ``an interstellar wind'' that flows past the Sun.
Within the heliosphere, ISN H is collision-less and the atoms can be regarded as individual particles following trajectories governed by solar forces until they are ionized and become pickup ions in the solar wind.

The Sun emits ultraviolet radiation, with the \lya line dominating the EUV part of the spectrum.
The \lya radiation resonantly excites the atoms by resonance scattering mechanism.
Within this mechanism, a photon is absorbed from the solar beam and subsequently re-emitted in a random direction. By using single scattering approximation, effectively those photons are removed from the solar beam and form the heliospheric backscatter glow, but for brevity we will still refer to this process as an absorption.
The absorption process involves a small instantaneous change in the atom momentum in the anti-solar direction.
Simultaneously, a photon is taken out from the solar photon flux and directionally redistributed within the heliosphere.
Thus, absorption of the solar EUV radiation on one hand is closely related to radiation pressure that the Sun exerts of H atoms, and on the other hand creates an absorption feature in the solar \lya line.

Absorption clearly reduces the solar \lya flux in the same spectral range that is responsible for solar radiation pressure.
Without absorption, the radiation pressure force drops with the square of distance, i.e., the ratio of radiation pressure to gravity is constant.
With absorption taken into account, this ratio will be distance-dependent. 

Absorption of solar radiation by ISN H inside the heliosphere has been pointed out in early heliospheric papers \citep[e.g.,][]{axford:72, meier:77a, wu_judge:79b} and more recently by \citet{IKL:18b} but it was largely neglected in the modeling of the distribution of ISN H inside the heliosphere, i.e., the atom dynamics has been treated assuming that the ratio of radiation pressure and solar gravity forces $\mu$ is independent on the location within the heliosphere.

Here, we assess the magnitude of absorption as a function of distance from the Sun and heliographic longitude and latitude.
We start with defining a theory of absorption of the solar \lya radiation by ISN H (Section \ref{sec:theory})
 We define an extension of the density, speed, and temperature of ISN H calculation scheme in the presence of absorption (Section \ref{sec:WTPMExt}).
With this, we investigate the spatial distribution of absorption in the heliosphere.
To that end, we define an iterative scheme of absorption features and density distribution (Section \ref{sec:calcScheme}), the calculation grid (Section \ref{sec:calcGrid}), and the initial conditions for the ISN H used in the simulations (Section \ref{sec:initial}).
We investigate the depth, width, and spectral offset of the heliospheric absorption feature in the solar \lya profile and parametrized it by the Gauss-like function (Section \ref{sec:params}).
The strength of absorption effect is measured by the attenuation factor $f_{abs}$ defined in Section \ref{sec:attenfact}.
We also investigate how a model density distribution inside the heliosphere changes when absorption (and thus radiation pressure) is taken into account (Section \ref{sec:Hdensity}).
In Section \ref{sec:solar_cycle} we show modulation of the absorption due to the solar activity cycle and in Section \ref{sec:TSdens} we discuss influence of the assumed density beyond the termination shock.
Finally, in Section \ref{sec:IBEXSignal} we show how absorption modifies the simulated flux of ISN H observed by Interstellar Boundary Explorer (IBEX). 

\section{Absorption - theory}
\label{sec:theory}
The theory of absorption of the solar \lya radiation by ISN H is adopted from a single scattering model presented by \citet{quemerais:06a}.
Following this approach, we  calculate the spectral irradiance $\text{I}(\myvec{r},\nu)$ at a given radius-vector $\myvec{r}$ and a frequency $\nu$ assuming that the initial spectral irradiance $\text{I}(\myvec{r_\text{E}},\nu)$ is that measured at $r_\text{E} = 1$ au, i.e., that ISN H is so rarefied inside 1 au that absorption in this region can be neglected.
This assumption is well fulfilled, as it is evident from observations of the solar \lya profile performed by \citet{lemaire_etal:15a} using the SOHO spacecraft, orbiting the Sun near the L1 Lagrange point.
In these observations, no absorption features are visible. 

In the approach originally followed by \citet{quemerais:06a}, the Sun is treated as an isotropic source of \lya radiation.
In the absorption theory we implement, we use a single-scattering approach, where photons after being absorbed by an atom (which is responsible for the absorption effect) are  treated as lost from the global photon population.
In the reality, they are re-emitted and form the helioglow. The re-emitted photons can potentially enter further absorption--re-emission sequences, which is referred to as the multiple-scattering effect.
Whether or not multiple-scattering is important for the heliosphere inside the termination shock, is still under discussion \citep[e.g., ][]{scherer_fahr:96, quemerais:00}.
Latest comparisons done by \citep{strumik_etal:21b} suggest that our WawHelioGlow model, that uses single scatter approach \citep{kubiak_etal:21a}, fits SOHO/SWAN measurements with a good agreement both in the downwind and the upwind directions.
Multiple-scattering is expected to modify the strongest the upwind to downwind helioglow intensity ratios for observations performed at 1 au \citep{quemerais:00}.
One of the possible explanations for the good agreement between our model and SWAN observations \citep[see][]{strumik_etal:21b} is that multiple scattering is not significant for helioglow observations at 1 au. Nevertheless, this requires further investigations, which we intend to perform in the future.
For the topic of this paper, multiple-scattering would only reduce the effect of absorption of direct solar radiation.

We calculate the spectral absorption following the formulae given by \citet{IKL:18b}, repeated below for readers' convenience. 
The spectral irradiance decreases with the square of solar distance because of purely-geometrical reasons and on top of that, there is another exponential reduction due to absorption that depends on the partial column density of the gas $n(\myvec{r})$ as well as on the cross section for absorption $\sigma_{cs}(\myvec{r},\nu)$, the thermal spread, and the radial component of the bulk velocity of ISN H in a given location in the heliosphere:

\begin{equation}
\label{eq:I_abs}
\frac{\text{I}(\myvec{r},\nu)}{\text{I}(\myvec{r_\text{E}},\nu)}=\left(\frac{r_\text{E}}{r}\right)^2\exp\left[-\left(\int^r_{r_\text{E}} n_{pr}(\myvec{r'})\sigma_{cs}(\myvec{r'},\nu,T_{g,pr}) dr'+\int^r_{r_\text{E}} n_{sc}(\myvec{r'})\sigma_{cs}(\myvec{r'},\nu,T_{g,sc}) dr'\right)\right],
\end{equation}

where $\text{I}(\myvec{r},\nu)$ is the spectral irradiance in the direction described by vector $\myvec{r}$; $\text{I}(\myvec{r_\text{E}},\nu)$ is the spectral irradiance measured at Earth's orbit (further on we will use our analytical model); $n_{pr}(\myvec{r})$ and $n_{sc}(\myvec{r})$ are the column densities of the primary and the secondary populations of ISN H, respectively;  $\sigma_{cs}(\myvec{r},\nu,T_{g,pr})$ and $\sigma_{cs}(\myvec{r},\nu,T_{g,sc})$ are the cross sections for absorption process calculated for the primary and the secondary populations of the ISN H.
Note that the cross section formula depends on the local temperature of the gas, which is different for each population.

The cross section is approximated as:
\begin{align}
\label{eq:cross_section1}
\sigma_{cs}(\nu)&=\sigma_0 \exp\left[-\left(\frac{\nu - \nu_0(1+u_r/c)}{\Delta \nu_D}\right)^2\right]\\
\label{eq:cross_section2}
\Delta \nu_D&=\frac{\nu_0}{c}\sqrt{\frac{2kT_g}{m_\text{H}}}\\
\label{eq:cross_section3}
\sigma_0&=\frac{\sigma_{tot}}{\sqrt{\pi}\Delta \nu_D},
\end{align}
where $\sigma_0$ is the cross section for $\nu=\nu_0$ (which is the case for atoms with null radial component of their velocity), $\Delta \nu_D$ is the Doppler width that is proportional to the thermal velocity of the gas at temperature $T_g$ at the distance $r'$, and $u_r$ is the radial component of the bulk velocity.
Other quantities used in our model along with their units in the cgs system are listed in Table \ref{tab:units}.
The cross section becomes Lorentzian when $\Delta \nu_D$ is much smaller than the natural width of the spectral line (which is true in the case of the \lya line).
Then it can be expressed as a Voigt function and easy integrated in an analytical way: 
\begin{align}
\label{eq:cross_section4}
\sigma_{tot}&=\int_0^\infty \sigma_{cs}(\nu)d\nu\\
&=\frac{\pi e^2}{m_e c}f_{osc}. \nonumber
\end{align}

\begin{deluxetable*}{llll}
\tablecaption{\label{tab:units} Relevant physical constants in the cgs units. In our numerical simulations, we use values based on the NIST standards.
}
\tablehead{\colhead{constant} & \colhead{symbol} & \colhead{value} & \colhead{unit}}
\startdata
elementary charge  & $e$   & $4.8032046729976595417 \times 10^{-10}$ & $ \text{cm}^{3/2} \text{ g}^{1/2} \text{ s}^{-1}$\\
electron mass     & $m_\text{e}$ & $9.1093837015 \times 10^{-28}$ & g \\
speed of light    & $c$     & $2.99792458 \times 10^{10}$  & $ \text{cm s}^{-1}$\\
\lya wavelength & $\lambda_0$ & $ 121.56701$ & nm\\
\lya frequency & $\nu_0$ & $ 2.46606755 \times 10^{15}$ & Hz\\
astronomical unit & $r_\text{E}$ & $1.495978707 \times 10^{13}$ & cm\\
hydrogen mass & $m_\text{H}$ & $1.6735328 \times 10^{-24}$ & g\\
H oscillator strength & $f_{osc}$ & $0.41641$ & dimensionless\\
Boltzmann constant & $k$ & $1.38064852 \times 10^{-16}$ & erg K$^{-1}$\\
total cross section & $\sigma_{tot}$ & $1.11 \times 10^{-2}$ & cm$^2$ s$^{-1}$\\
gravity const. $\times$ solar mass & GM$_\sun$ & $1.327124421864553 \times 10^{26}$ & cm$^3$ s$^{-2}$\\
\enddata
\end{deluxetable*}

Later in this paper we will be using a dimensionless factor $\mu$, which is defined as a ratio between the force caused by the momentum transfer due to photon scattering events and gravitational force (see Equation \ref{eq:radpress}).
When $\mu = 1$, there is no effective force acting on a moving atom. 

\begin{align}
\label{eq:radpress}
\nonumber
\mu&=\frac{\text{P}_\text{rad}}{|\text{F}_\text{g}|}\\ \nonumber
&= \text{I}(\myvec{r},\nu_0) \frac{\pi \text{e}^2}{\text{m}_e \text{c}}\frac{\text{h}\lambda_0}{\text{c}}\text{f}_{\text{osc}} \frac{\text{r}_\text{E}^2}{\text{GM}_\sun\text{m}_\text{H}}\\
&=\frac{\text{I}(\myvec{r},\nu_0)}{\text{p}_\text{H}},\\
\label{eq:ph}
\text{p}_\text{H} & = \left[\frac{\pi\, \text{e}^2}{\text{m}_\text{e}\,c}\,\frac{\text{h}\lambda_0}{\text{c}} \text{f}_{\text{osc}} \frac{r_\text{E}^2}{\text{G}\,\text{M}_\sun \text{m}_\text{H}}\right]^{-1} = 3.34467\times 10^{12} & \text{ ph s}^{-1}\,\text{cm}^{-2}\,\text{nm}^{-1}.
\end{align}
More details can be found in Appendix in \citet{kubiak_etal:21a}.
The advantage of using the factor $\mu$ is that it is independent from the distance when absorption is neglected.
Also we will often refer to the radial velocity ($u_r$) of the ISN H atoms rather than their resonance frequency.
According to the Doppler law, there is an easy way to translate one into the other:
\begin{equation}
u_r=c\left(1-\frac{\nu_0}{\nu}\right)=-c \frac{\Delta \lambda}{\lambda_0},
\label{eq:ur}
\end{equation}
where $u_r$ is the radial component of the ISN H atom velocity, $\nu$ is the shifted frequency corresponding to the velocity $u_r$, and $\nu_0$ is listed in Table \ref{tab:units}.

\section{Extension of the WTPM code to accommodate absorption}
\label{sec:WTPMExt}
The Warsaw Test Particle Model \citep[WTPM,][]{tarnopolski_bzowski:09} of ISN H distribution inside the termination shock is based on the paradigm of the so-called hot model \citep{fahr:78, fahr:79, thomas:78, wu_judge:79a}.
The hot model paradigm is based on a solution of the Boltzmann equation, which governs the spatial distribution of a collision-less gas streaming past the Sun and subjected to ionization losses, by method of characteristics.
The density is calculated by numerical integration of the local distribution function of the gas, which is computed as composed of a number of characteristics, modeled as trajectories of individual test particles, corresponding to ISN H atoms.
The goal is to relate the magnitude of the distribution function at the location inside the heliosphere to its value far away from the Sun, in the so-called source region, where the distribution function of the ISN gas is assumed to be known.
The relation is by the Liouville theorem, with the kinematic parameters modified by the force acting on the atoms, and the magnitude of the local distribution function modified by losses due to ionization. 

Assuming that the force is central, constant in time, and decreasing with the square of the distance from the Sun, that there is no absorption of the solar radiation within the ISN gas, and that the ionization rate also decreases with the distance squared, the equation of motion of a H atom is given by 
\begin{equation}
\frac{d^2 \myvec{r}(t)}{d t^2} = -\frac{G M_\sun (1 - \mu)}{r(t)^2}\frac{\myvec{r}(t)}{r(t)},
\label{eq:eqMotionKepler}
\end{equation}
where $\myvec{r}(t)$ is the radius vector of a H atom at a time $t$, $r = |\myvec{r}|$, $G$ is the gravitational constant, $M_\sun$ the solar mass, and $\mu$ is a factor of compensation of the solar gravity force by the solar radiation pressure force.
The objective of this equation is to provide a relation between the velocity vector of the atom $\myvec{u}_0 = d \myvec{r}_0/dt$ at a location inside the heliosphere given by the radius-vector $\myvec{r}_0$ with the velocity vector $\myvec{u}_{\text{ISM}}$ in the source region.
When $\mu$ is a constant, one does not need to solve the equation of motion explicitly.
Instead, $\myvec{u}_0$ can be related to $\myvec{u}_{\text{ISM}}$ by the hyperbolic Kepler equation.
This is true regardless of $\mu < 1$ (for an effective attractive force) or $\mu > 1$ (for an effective repulsive force).
Out of two possible solutions, one needs to select that corresponding to an earlier time, i.e., the equation of motion must be solved backward in time. 

In the reality, however, both the ionization rate and the magnitude of the force acting on H atoms inside the heliosphere vary with time because of the changes in the solar activity.
This causes the solar radiation pressure $\mu$ to become a function of time: $\mu \equiv \mu(t)$, but the radiation pressure still drops with the square of solar distance and thus can be expressed as a certain time-dependent factor compensating the solar gravity force.
To address the situation when the $\mu$ factor is time dependent but does not vary with heliolatitude, \citet{rucinski_bzowski:95b} suggested to use a numerical scheme to solve the equation of motion all the way from $\myvec{r}_0$ to the source region to provide $\myvec{u}_{\text{ISM}}$ along with the attenuation factor due to ionization losses \citep[see Equations 2 and 3 in][]{bzowski_etal:13b}.
In this case, the equation of motion \ref{eq:eqMotionKepler} with $\mu \equiv \mu(t)$ can be reduced to a two-dimensional form.

\citet{tarnopolski_bzowski:09} realized that the solar spectral flux varies significantly within the \lya line and that a H atom in its solar orbit is sensitive to different portions of this profile due to the Doppler effect, depending on its radial velocity $u_r(t) = \myvec{u}(t)\cdot \myvec{r}(t)$.
They also pointed out that the solar \lya profile shape (the spectral flux) $I_\nu(\nu)$ varies depending on the total instantaneous flux within the \lya line $I\tot = \int I_\nu(\nu)\, d\nu$.
Since the radiation pressure factor is proportional to the solar spectral flux, and the atom's radial velocity is related by the Doppler law to the wave frequency $\nu$, the radiation pressure factor $\mu$ in Equation \ref{eq:eqMotionKepler} becomes a complex function of radial velocity and the total solar flux in the \lya line, which varies with time and, as recently demonstrated by \citet{strumik_etal:21b}, with heliolatitude $\phi$: $\mu \equiv \mu(u_r, I\tot(t, \phi))$.
Still, the magnitude of radiation pressure force drops with the square of solar distance.
While the motion of individual atoms is planar because the force is central, the equation of motion is solved numerically and it is more convenient to use its three-dimensional form.
The dependence of $I_\text{tot}$ on heliolatitude is approximated by an analytic function \citep{bzowski_etal:13a}, and its dependence on time is obtained from in-ecliptic measurements \citep{machol_etal:19a} and tabulated on a fixed time grid with the nodes at centers of Carrington rotation periods. The dependence of $\mu$ on $u_r$ for a an $I\tot$ value characteristic for a time moment $t$ was given by \citet{IKL:20a}. 

The WTPM model with these considerations implemented in the radiation pressure model was presented by \citet{kubiak_etal:21a}.
But \citet{IKL:18b} assessed the magnitude of absorption of the solar \lya radiation by the ISN H gas inside the heliosphere and pointed out that absorption of the solar radiation on ISN H results in a more rapid drop of radiation pressure with the solar distance than $1/r^2$. In this paper, we modify the WTPM model to take this effect into account. 

Absorption adds an additional narrow line to the solar line profile.
As discussed later in the paper (see Equation \ref{eq_mu}), this feature can be represented as a certain factor $1- F_{fit}(\myvec{r}, u_r)$ attenuating the radiation pressure factor $\mu$.
The magnitude of $F_{fit}$ varies with the distance and the location within the heliosphere.
In the WTPM, it has a form of an analytic function (see Equation \ref{eq:fit_fun}) with the parameters calculated on a fixed grid in the 3D space.
Between the grid points the absorption profile parameters are linearly interpolated during the calculations.
Thus, the radiation pressure factor $\mu$ is now a function of all three coordinates in the heliographic reference system and the equation of motion of a H atom becomes:
\begin{equation}
\frac{d^2 \myvec{r}(t)}{d t^2} = -\frac{G M_\sun (1 - \mu_{abs})}{r(t)^2}\frac{\myvec{r}(t)}{r(t)},
\label{eq:eqMotionAbsorption}
\end{equation}
with 
\begin{equation}
\mu_{abs} \equiv \mu\left(u_r,I_\text{tot}(t) I_{\lambda \phi}(\phi)\right)\,\left(1 - F_{fit}(\myvec{r}, u_r)\right),
\label{eq:muEff}
\end{equation}
and $\mu_{abs}$ varies both with the distance from the Sun and the direction of the radius-vector $\myvec{r}$.
In Equation \ref{eq:muEff}, $\mu(u_r,I\tot(t))$ corresponds to the radiation pressure acting on a H atom at 1 au from the Sun in the solar equator plane.
When this atom has a radial velocity $u_r$ at a time $t$, then the magnitude of the total solar \lya flux at solar equator is equal to $I\tot(t)$.
This function $\mu(u_r, I\tot(t))$ is defined by \citet{IKL:20a}.
The factor $I_{\lambda\phi}(\phi)$ describes the variation of $I\tot$ with heliolatitude $\phi$, as discussed by \citet{kubiak_etal:21a}.
Hence, away from the solar equator at a heliolatitude $\phi$ but still at 1 au, the radiation pressure is calculated from the formula given by \citet{IKL:20a} parametrized by the total solar \lya flux modulated by the adopted heliolatitude dependence, equal to $I\tot(t) I_{\lambda \phi}(\phi)$.
In the present version, $I_{\lambda\phi}(\phi)$ is assumed to be invariable with time, but this can be easily changed to accommodate a time variation of this parameter, which might be needed according to \citet{strumik_etal:21b}.

The drop of radiation pressure with the solar distance is described by the absorption feature $1 - F_{fit}(\myvec{r}, u_r)$, given by Equation \ref{eq:fit_fun}.
Note that with these definitions, we can still use the radiation pressure compensation function $\mu_{abs}$, even though the radiation pressure force now decreases more rapidly with the distance than $1/r^2$. Also note that still, the total force acting on H atoms is central in the solar-inertial frame.

\section{Simulations}
\label{sec:calculations}
\noindent
\subsection{Simulations scheme}
\label{sec:calcScheme}
The simulations were performed in the following steps:
\begin{enumerate}
\item Baseline calculations of the density of ISN H with no absorption, where the density and bulk velocity distributions of ISN H were calculated using the nWTPM code in the grid points for the selected epochs.
\item Calculation of absorption profiles in all nodes of the grid.
\item Approximation of the absorption profiles using analytic functions (see Section \ref{sec:params}).
\item Adoption of a modified model of radiation pressure for all grid points, with absorption effects taken into account.
\item Calculation of the density distribution of ISN H using the appropriately modified nWTPM model with the absorption-modified radial velocity-dependent radiation pressure. 
\item Iterating absorption estimates--density calculations until the difference in the densities between subsequent iterations became smaller than 0.5\% (it turned out that this criterion was satisfied after two iterations).  
\end{enumerate}

\subsection{Simulation grid}
\label{sec:calcGrid}
The calculations were performed on a 3D spatial grid in the heliographic coordinates, centered at the Sun. The nodes are separated by 10\degr{} in heliographic longitude and latitude.
The radial nodes are distributed from r=0.1 au to r=140 au.
The outer boundary is set to the location of the TS in the downwind direction.
Our model is intended to be used inside the TS, so depending on preferred global model of the heliosphere, some of the grid points should be excluded from discussion.
In this paper, we present all results up to the distance of 140 au in each direction, so the reader can decide where the calculations should be stopped due to the TS position.
The grid points are shown in Figure \ref{fig:params}.

\subsection{Initial conditions}
\label{sec:initial}
Following \citet{IKL:18b}, we assumed that ISN H inside the heliosphere is a superposition of the primary and secondary populations with the densities adding up to 0.085~cm$^{-3}$ on the boundary of our calculations (set at 300 au), based on \citet{bzowski_etal:09a}.
The inflow direction of the primary population in the heliographic coordinates are given by lon=179.35\degr{} and lat=5.13\degr{} (which corresponds to the ecliptic coordinates: lon=255.75\degr{} and lat=5.17\degr{}), while the inflow direction of the secondary population is consistent with the so called Warm Breeze \citep{kubiak_etal:16a}: lon=251.57\degr{} and lat=11.95\degr{} in the J2000 ecliptic coordinate system.
Since recently \citet{swaczyna_etal:20a} suggested that the density at the TS might be, in fact, larger by $\sim 45$\%, we repeated some of the simulations for the densities of the two populations appropriately scaled up, with the temperatures and bulk velocities of the two populations unchanged (see Section \ref{sec:TSdens}).

The simulations of the gas distribution were performed using the nWTPM model of ISN H \citep{tarnopolski_bzowski:09} \citep[see also][]{kubiak_etal:21a} with the ionization rate being a function of time and heliolatitude following \citet{sokol_etal:20a}.
The radiation pressure at 1 au was taken from \citet{IKL:20a}.
Since the solar output varies during the solar cycle, we performed additional simulations for high and low solar activity (see Section \ref{sec:solar_cycle}).
The simulations were performed for three epochs during the solar activity cycle: 1996 (low activity), 2002 (high activity), and 1999 (medium activity). The baseline simulations are for medium activity, and the high and low activity cases are used to assess the time dependence of absorption effects.

\subsection{Inclusion of absorption}
When calculating the density using the nWTPM, we track the trajectories of H atoms in all sky directions for all speeds.
To obtain the density, the results of tracking of individual trajectories, which correspond to characteristics of the Boltzmann equation, are integrated over speed, which yields the partial density in a given location in space.
Subsequently, this partial density is integrated over the directions covering the entire sphere.
When an atom is tracked numerically on its trajectory, in each point of the trajectory we need to have an exact radiation pressure dependent on time, to calculate the equation of motion. 

The dependence of radiation pressure on time was first introduced by \citet{rucinski_bzowski:95b}.
These authors assumed that the radiation pressure force was a distance-independent fraction of the solar gravity force, and this assumption was maintained in the WTPM code until now. In the approach from the present paper, the radiation pressure in each point and time moment is reduced by the absorption.
 
First, we calculate the density of hydrogen on the heliolatitudinal grid, with no absorption included. Then, for each point of our grid we estimate the four parameters of absorption: $A_a$, $\xi_a$, $\sigma_a$, and $n_a$  (see Section \ref{sec:params}).
We prepare four files with these parameters set on our 3D grid.
Now we can use them to re-calculate the density.
For each atom tracked, for each point of its trajectory, we calculate the effective radiation pressure with absorption included according to Equation \ref{eq:muEff}.
Finally, we use our new $\mu_{abs}$ to calculate numerically the trajectories of hydrogen atoms, to integrate over them to obtain the local density and the local velocity along with other parameters of the ISN H in the grid points.

\section{Results}
\label{sec:Results}
\subsection{Parameters of the absorption profiles}
\label{sec:params}

As a result of the simulations presented in Section \ref{sec:calculations}, in each node of the computation grid we have a full \lya profile as a function of radial velocity (equivalently to frequency or wavelength), with the absorption feature.

\begin{figure*}[ht!]
\centering
\includegraphics[width=0.7\linewidth]{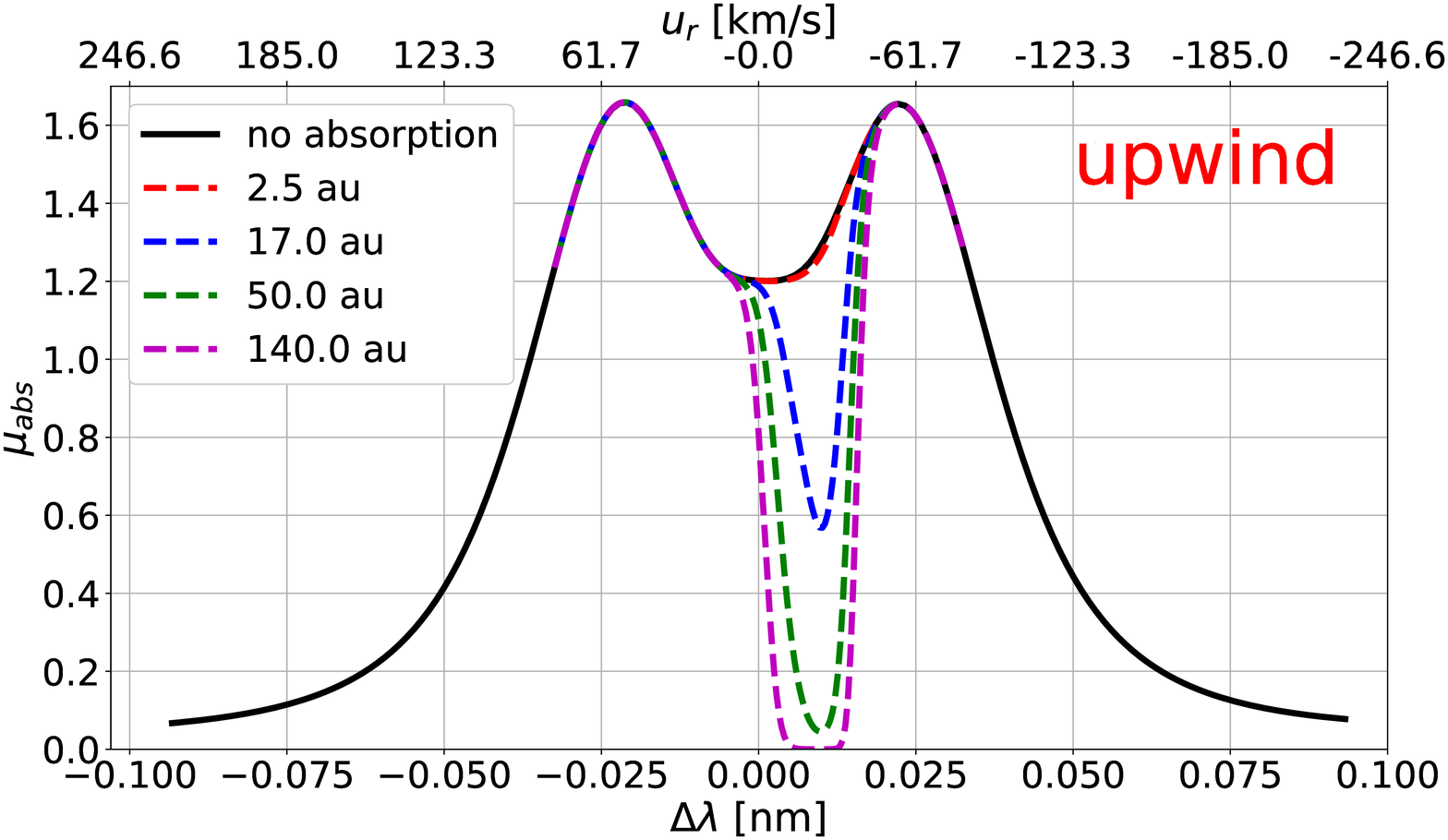}
\includegraphics[width=0.7\linewidth]{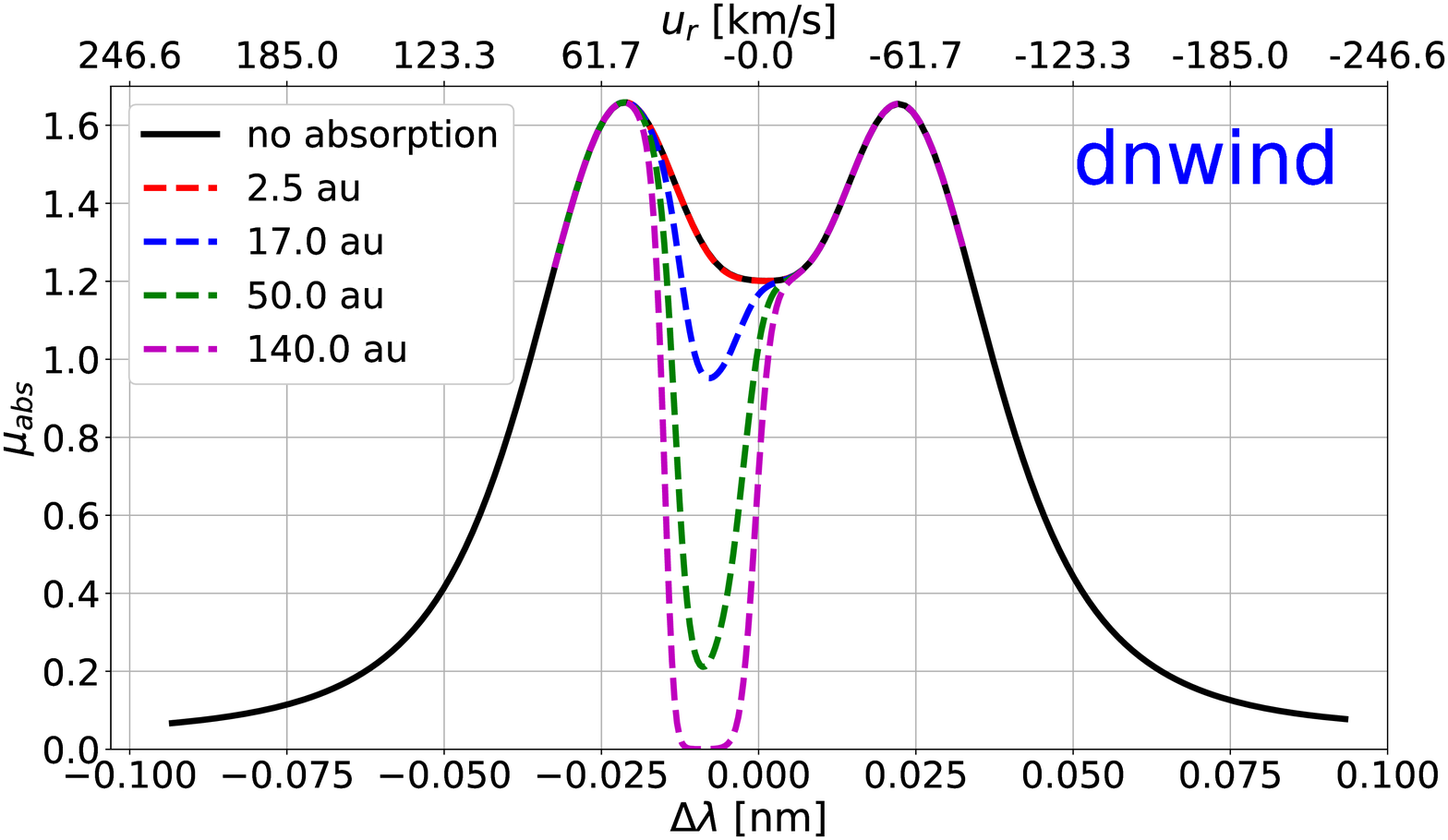}
\caption{The radiation pressure
parameter $\mu$ as a function of radial velocity, including absorption
effects, found for the solar \lya{} profile given by  Equation \ref{eq:radpress} and
corrected for absorption with Equation \ref{eq:muEff}.
Black solid line shows the original profile without absorption, while dashed lines correspond to profiles at various distances from the Sun (red -- 2.5 au, blue -- 17 au, green -- 50 au, and magenta -- 140 au) with absorption included. The top panel presents the upwind direction, while the bottom panel the downwind direction. The profiles are calculated for the medium solar activity in 1999.0.}
\label{fig:absorb_prof}
\end{figure*}

Figure \ref{fig:absorb_prof} shows several examples of absorbed profiles seen at different distances in two directions: upwind and downwind. In the upwind direction, our profile is absorbed on the right side (for $\Delta \lambda>0$). The atoms that are coming from the upwind
direction towards the Sun have negative radial velocities. From the H atom point of view, the source of the photons is approaching, so the wavelength of these photons is shifted towards shorter values
due to the Doppler effect. It means that in the atom rest frame photons that originally have longer
wavelengths than \lya are shifted, and if they end up at 121.567 nm, they will be absorbed.
Therefore, the right side of the original profile will be affected. The situation is reversed in the downwind direction, where the left side of the profile is affected. 
The depth of the absorbed part of the profile depends on the column density in the given direction.
We can compare Figure \ref{fig:absorb_prof} with the Figure 3 from \citet{wu_judge:79a}, where absorption is shown to affect the opposite side of the profile. It is a result of a mistake in their formula, which was corrected by the authors in erratum published later.
The absorption effect in \citet{wu_judge:79a} is slightly stronger than in our paper. There could be few reasons: we assumed a smaller density of the ISN H beyond the heliosphere, but also we have a much more sophisticated model of radiation pressure and ionization that depends on the solar activity phase. Qualitatively, however, the absorption features predicted by \citet{wu_judge:79a} and by us are similar.

\begin{figure*}[ht!]
\centering
\includegraphics[width=0.31\linewidth]{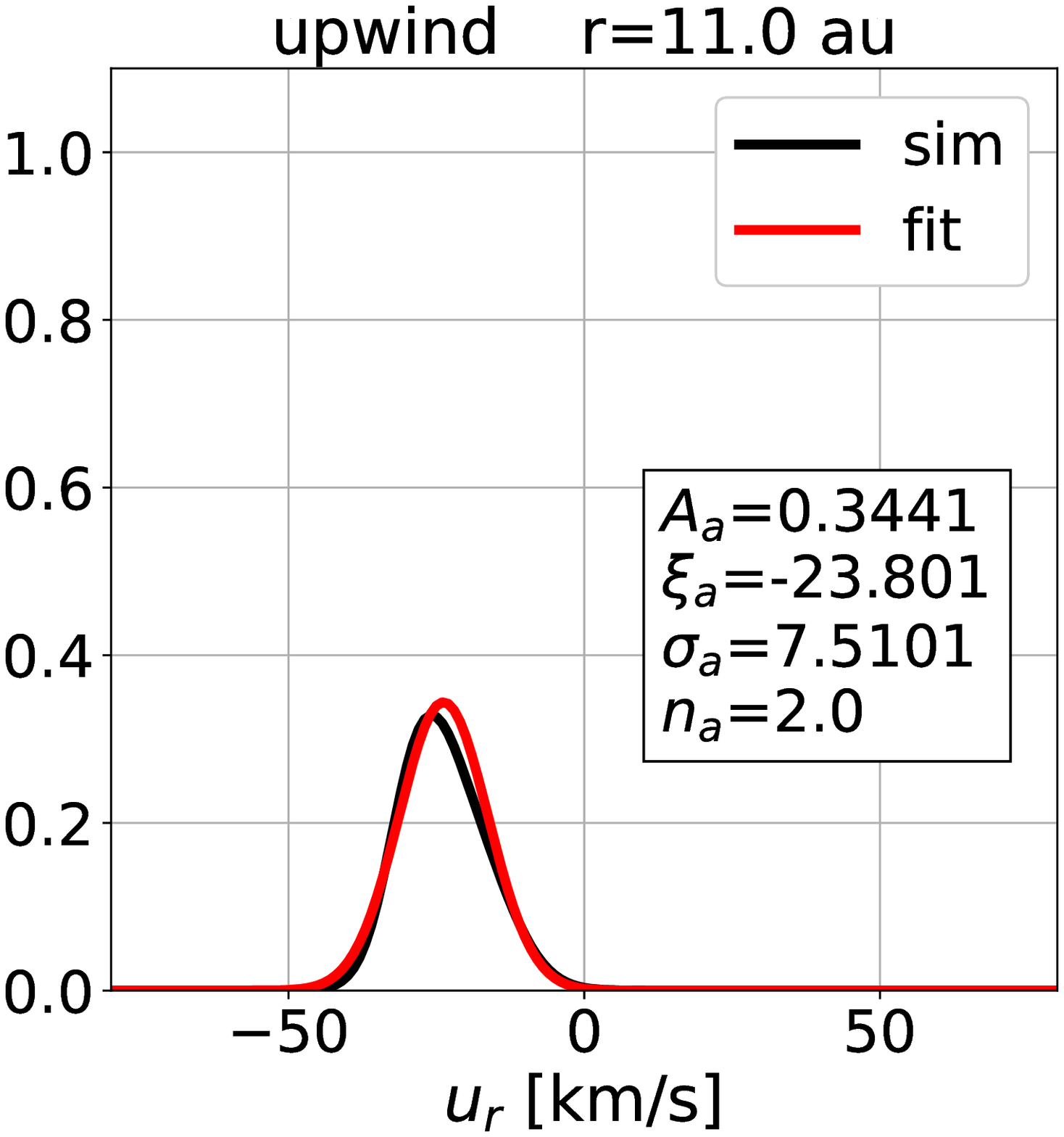}
\includegraphics[width=0.31\linewidth]{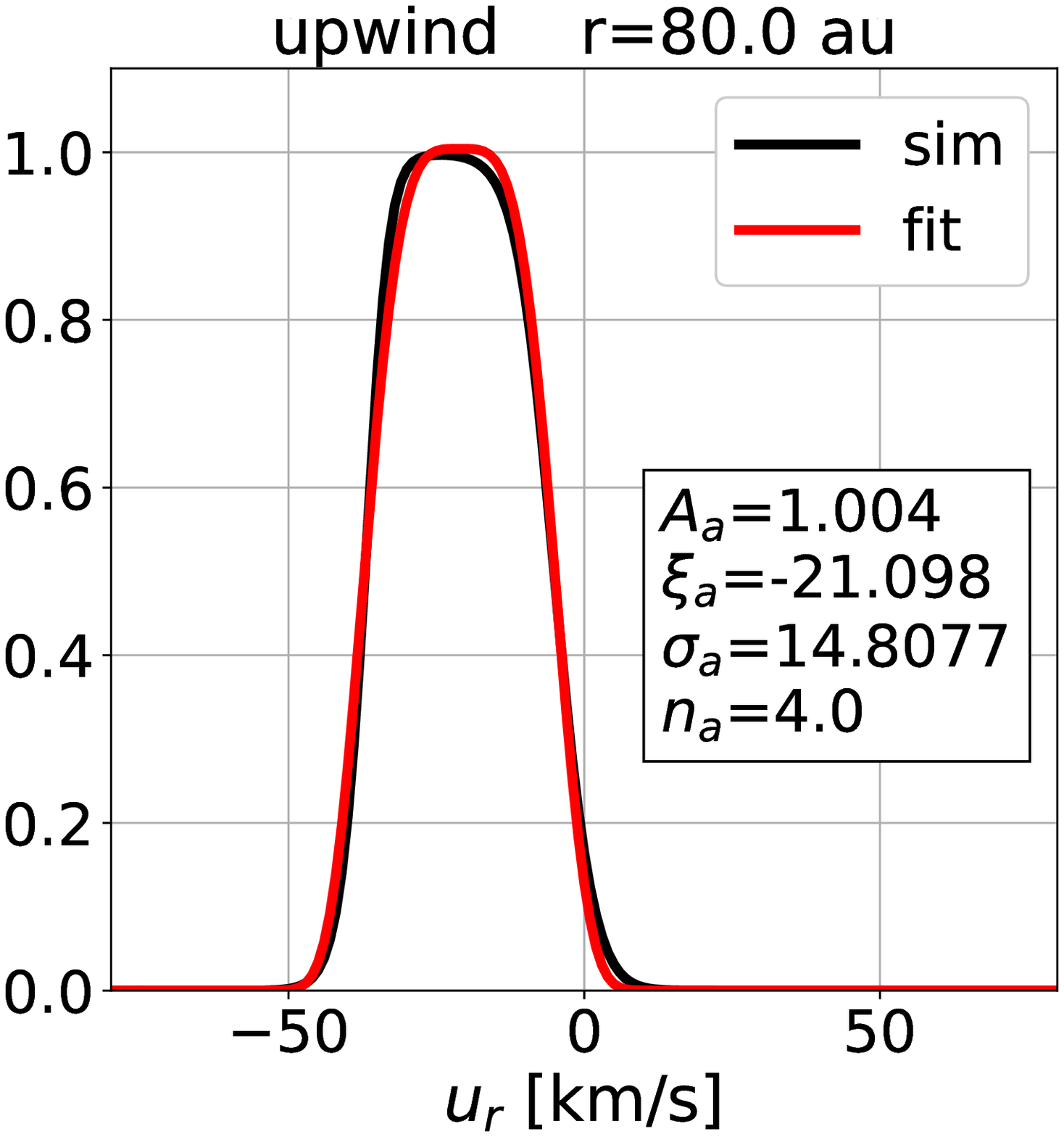}
\includegraphics[width=0.31\linewidth]{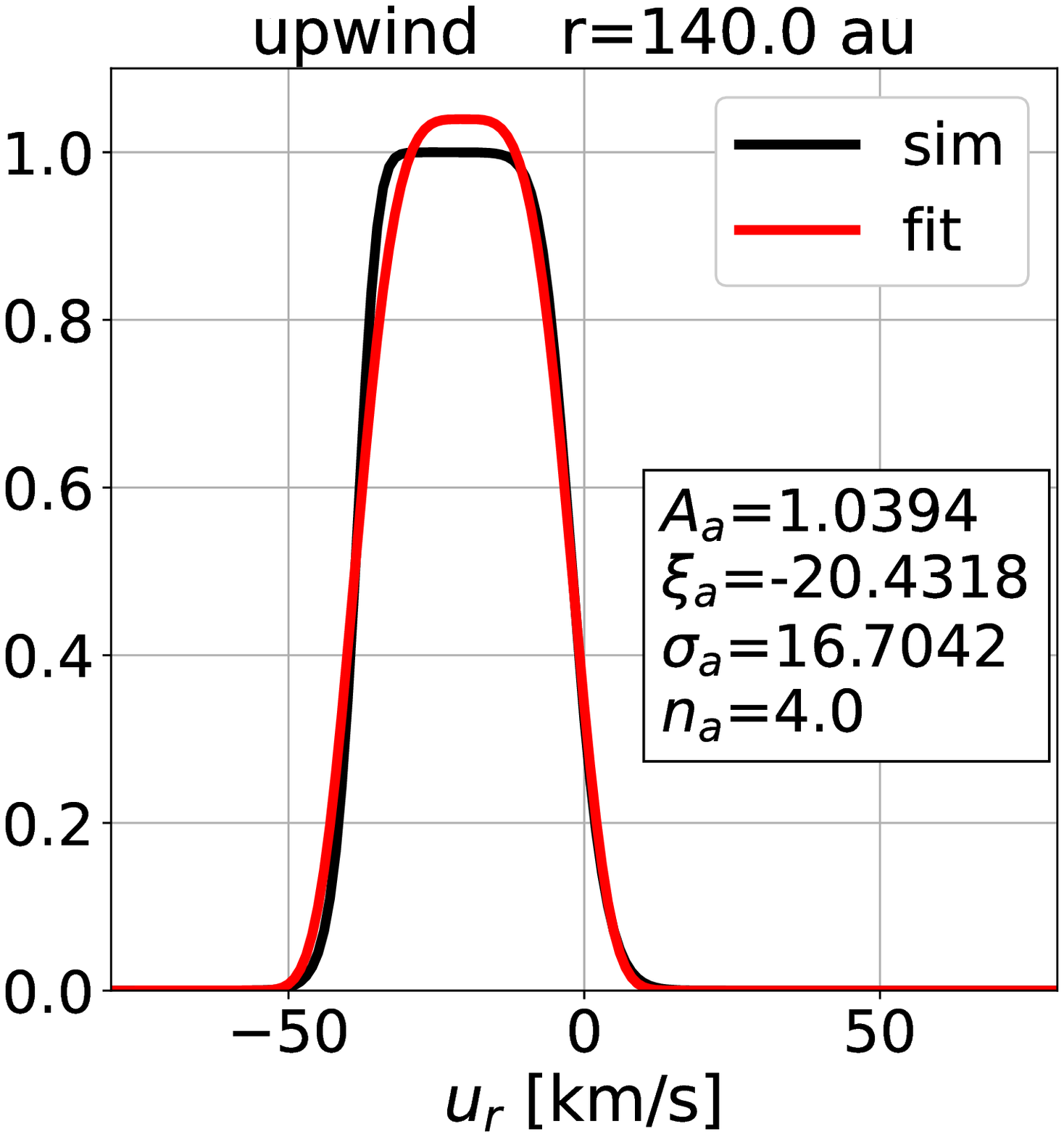}
\caption{Example absorption profiles calculated according to Equation \ref{eq:I_abs} (black line) and approximation of the absorption profiles according to Equation \ref{eq:fit_fun} (red line). All of them showing the absorption effect in the upwind direction at three distances: 11 au -- left side panel, 80 au -- center panel, 140 au -- right side panel.}
\label{fig:fit_fun}
\end{figure*}

Had the ISN gas been homogeneous, the absorption profiles would be Gaussian-like.
In the reality, however, the gas along any radial line features a gradient in radial velocity as well as a gradient in the temperature.
In addition, the gas is a superposition of two populations, with different densities, flow directions, speeds, and temperatures.
As a result, the absorption profiles become non-Gaussian, as illustrated by black lines in Figure \ref{fig:fit_fun}.
Since on one hand, full 3D time dependent simulations of the absorption effect are very time consuming, and on the other hand the effect of absorption in the radiation pressure model should be relatively easy to implement, we decided to introduce approximation formulae to model the spectral irradiance with absorption included. 

Absorption may result in saturation of the absorption feature in the solar \lya line.
The saturation starts at the center of the absorption feature and progresses towards longer and shorter wavelength with the increase of the solar distance.
Before the onset of saturation, the shape of the absorption feature can be approximated by the Gaussian function.
With the onset of saturation, we change the approximation function by allowing the parameter $n_a$ to be equal to 4 (instead of $n_a=2$ in the case of the Gaussian function) in Equation \ref{eq:fit_fun}. 
\begin{align}
\label{eq:fit_fun}
F_{fit}(u_r)&=A_a \exp\left[- \frac{1}{2}\left(\frac{u_r - \xi_a}{\sigma_a}\right)^{n_a}\right]
\end{align}
The transition between one function and the other occurs at different distances and depends on the selected direction in space.

Using this procedure instead of the full absorption profile at each grid point, we obtain a set $\myvec{q} = \{A_a, \xi_a, \sigma_a, n_a \}$ of four parameters of approximate absorption profiles: the amplitude ($A_a$), the mean Doppler shift ($\xi_a$), the dispersion ($\sigma_a$), and the type of fitting function ($n_a$), which can be equal to 2 (the Gaussian function) or 4.
The values of these parameters for our computation grid are available as a additional table in MRT standard attached to this paper (see Table \ref{tab:params} as an example), as well as on the webpage\footnote{http://users.cbk.waw.pl/~ikowalska/index.php?content=abs}.
Along with the \lya radiation pressure model described in \citet{IKL:20a}, these data allow to calculate the modified radiation pressure that includes the absorption effect at any location within the calculation grid.

At greater distances, our fit returns amplitudes greater than 1 (see Figure \ref{fig:fit_fun}, right hand panel), which is non-physical, therefore in these parts of the profile we set $F_{fit}=1$ in our calculations.

\begin{figure*}[ht!]
\centering
\includegraphics[width=0.3\linewidth]{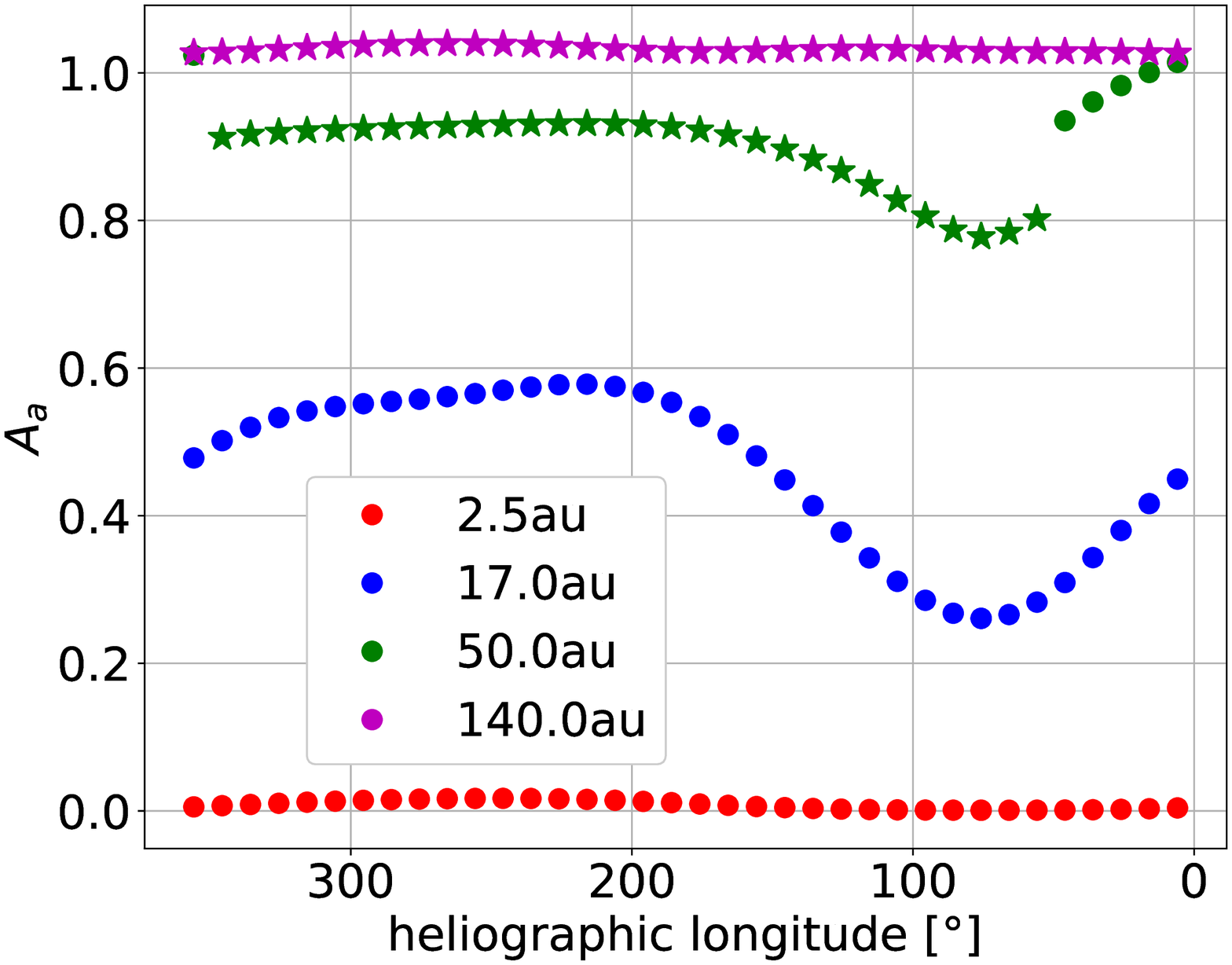}
\includegraphics[width=0.3\linewidth]{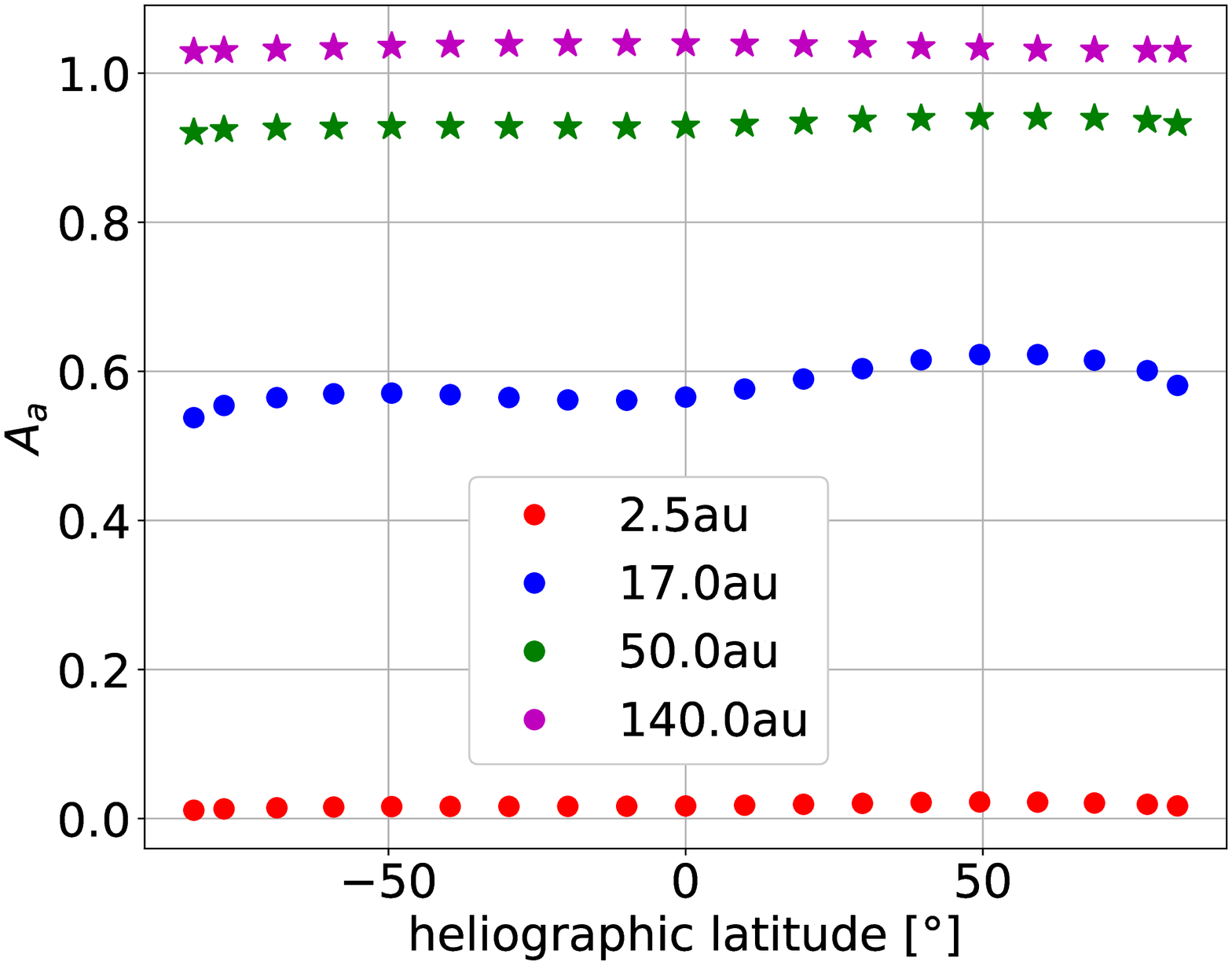}
\includegraphics[width=0.3\linewidth]{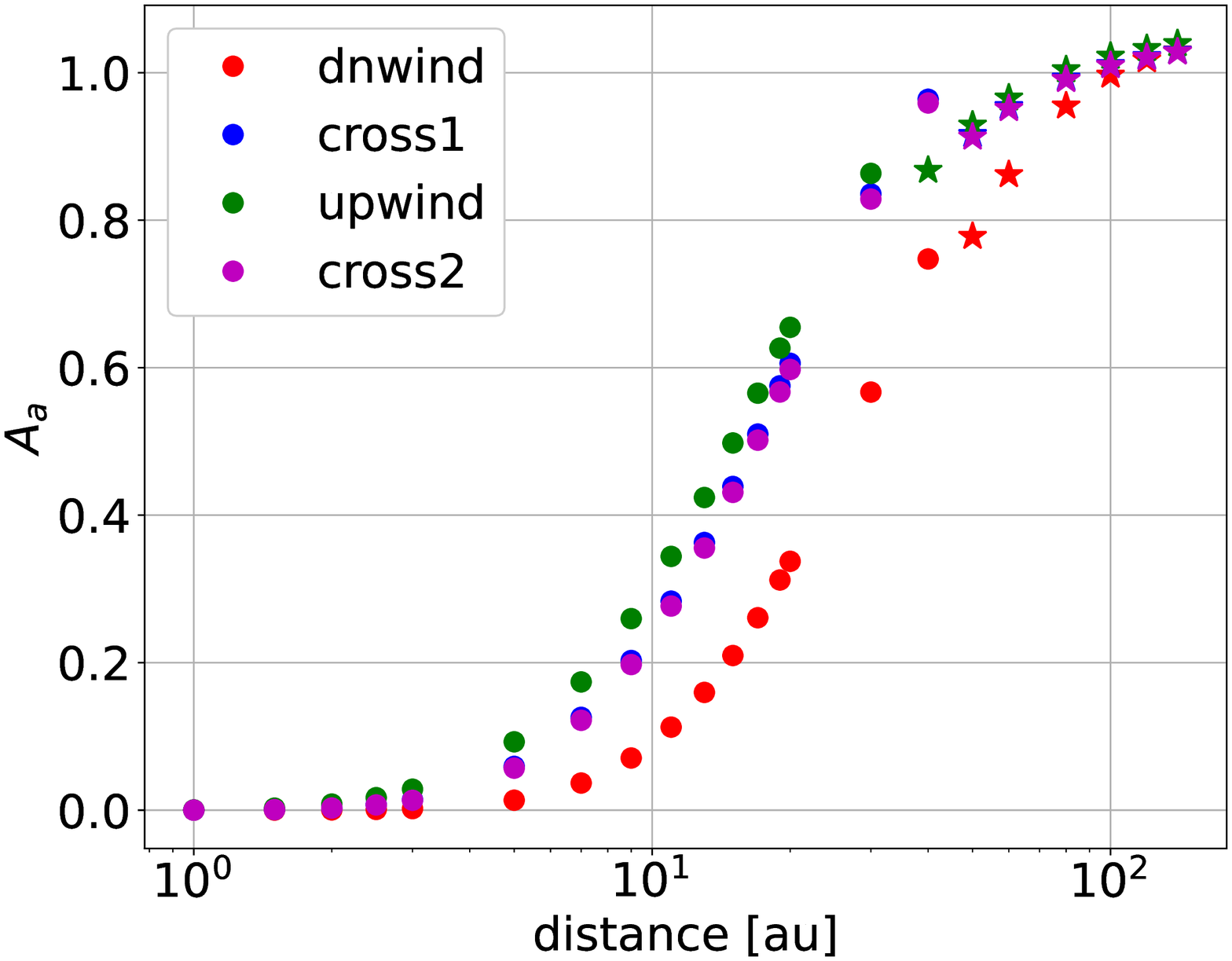}
\includegraphics[width=0.3\linewidth]{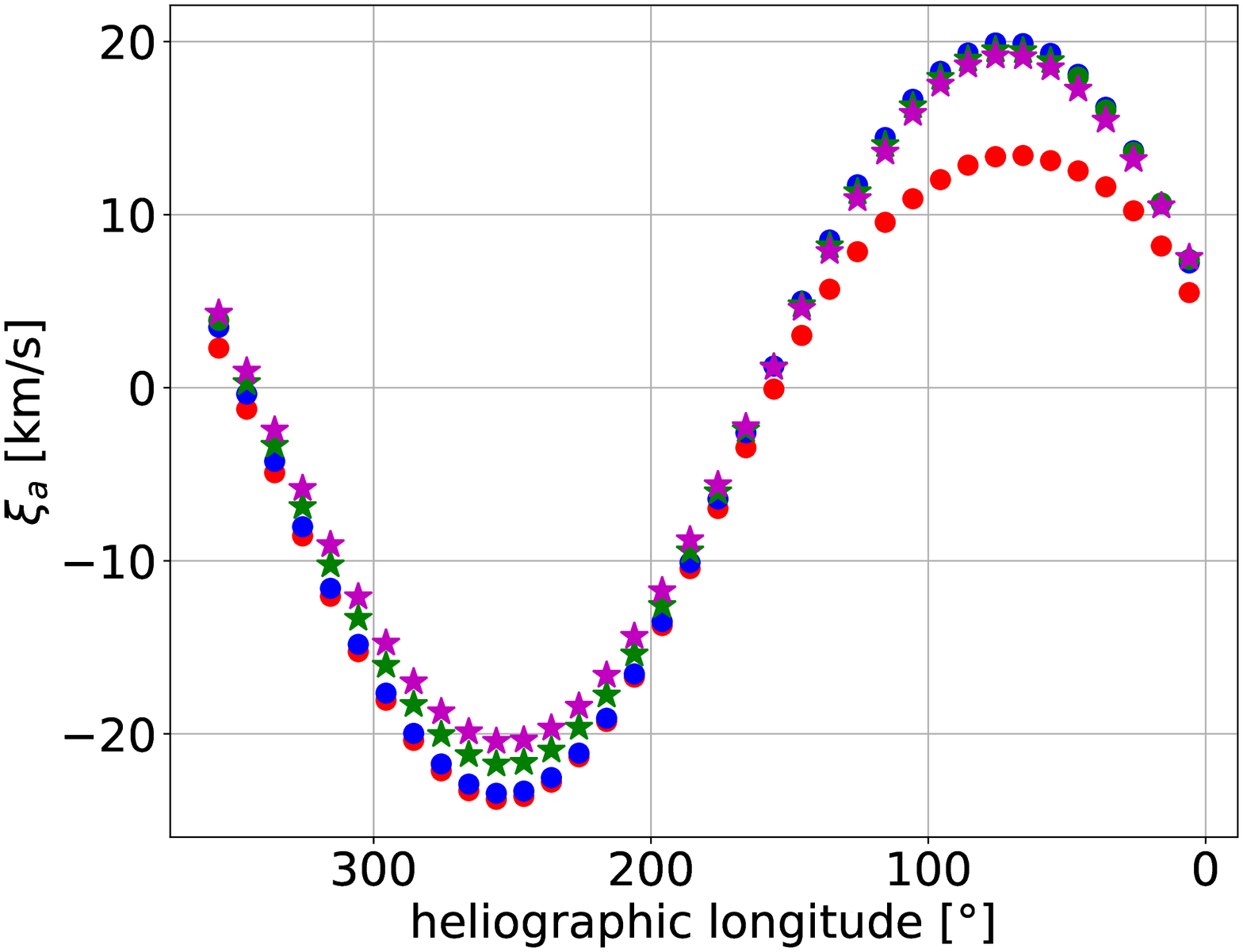}
\includegraphics[width=0.3\linewidth]{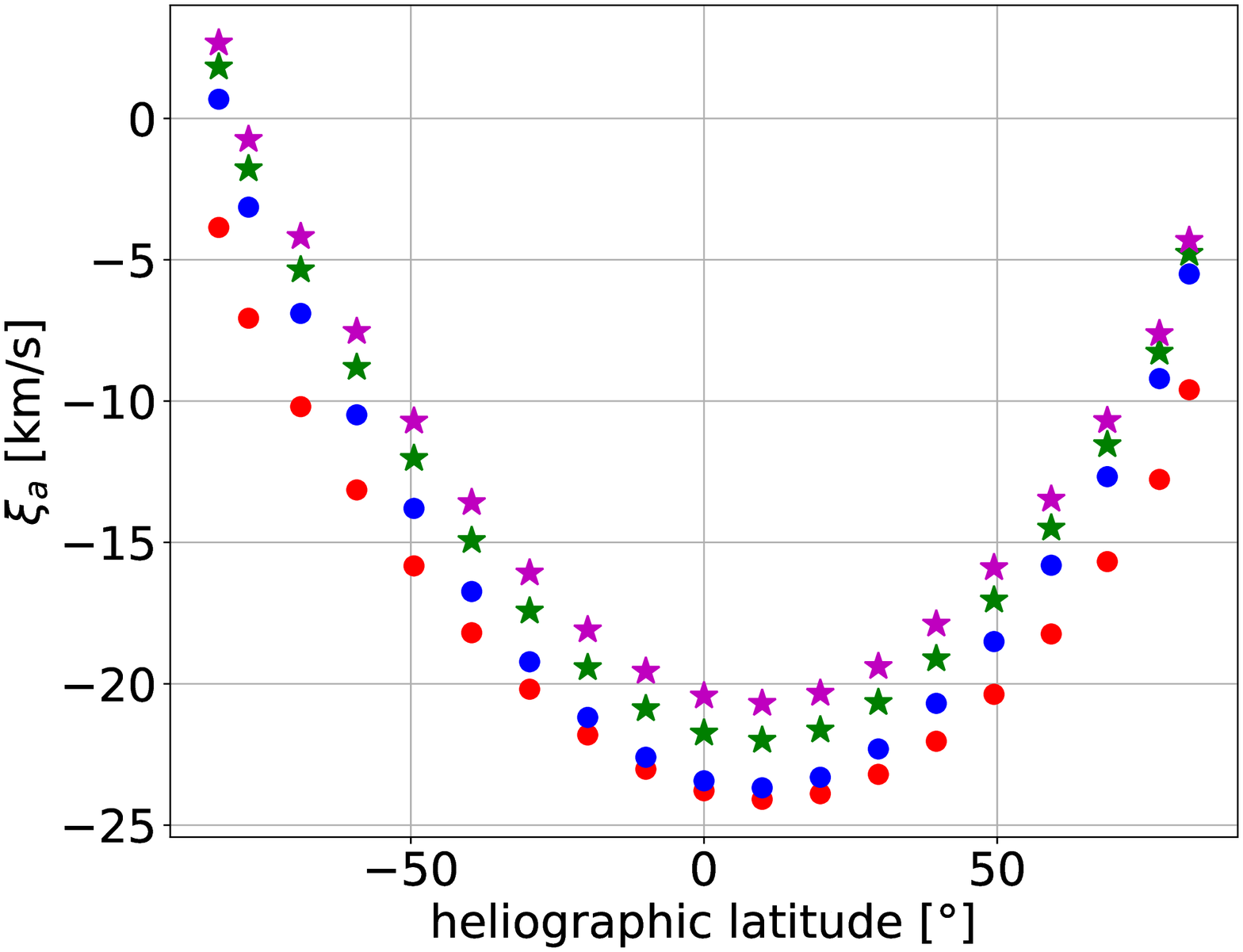}
\includegraphics[width=0.3\linewidth]{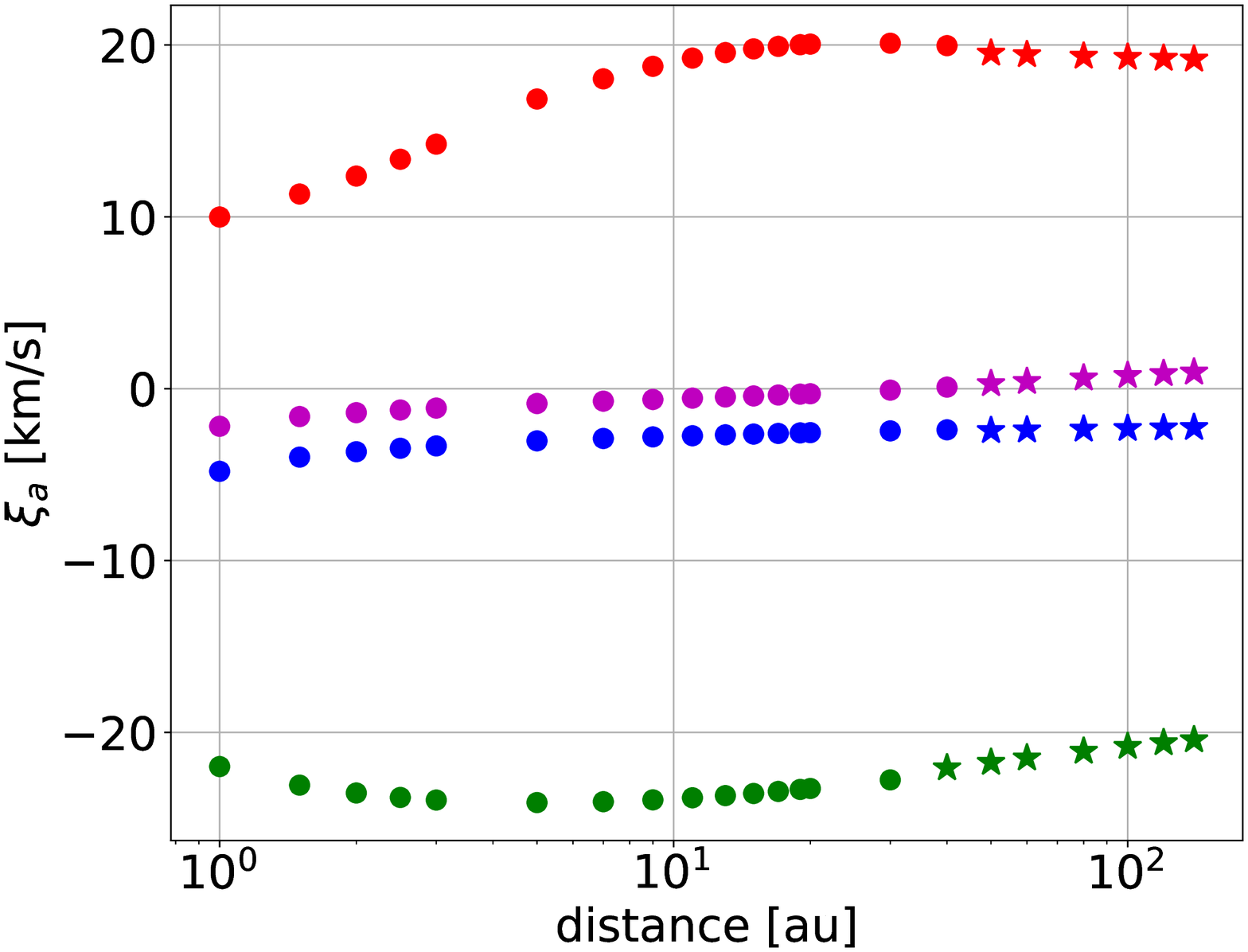}
\includegraphics[width=0.3\linewidth]{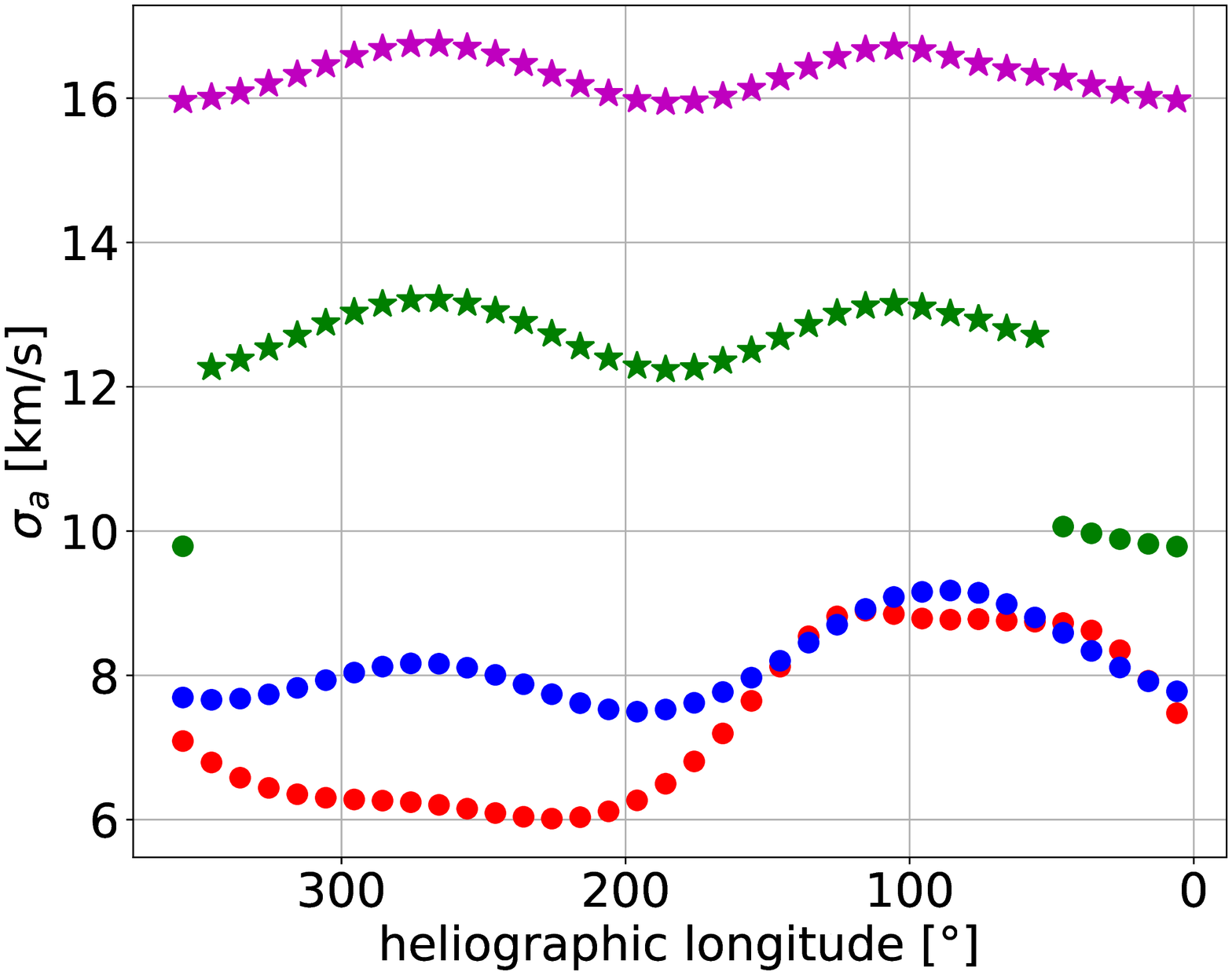}
\includegraphics[width=0.3\linewidth]{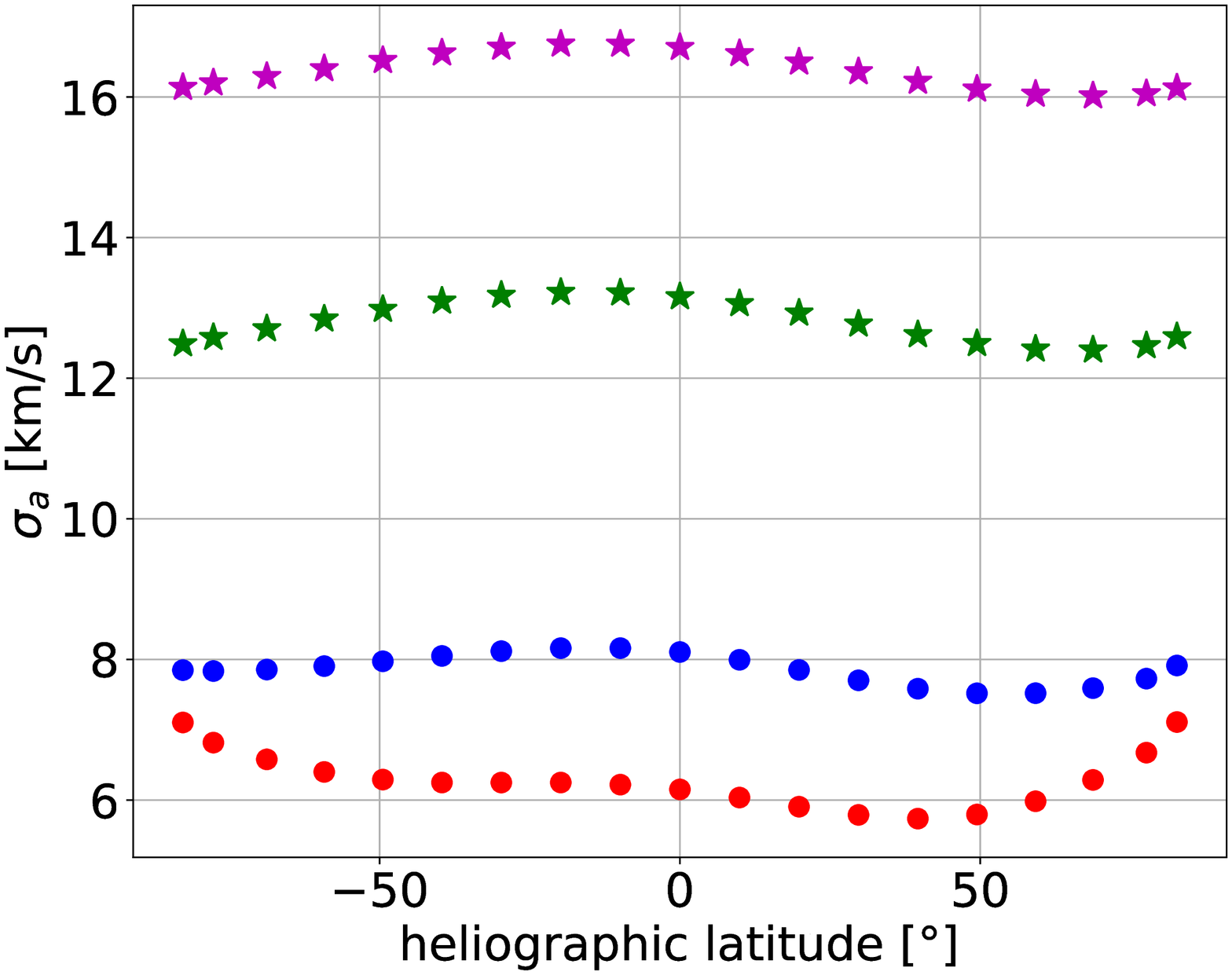}
\includegraphics[width=0.3\linewidth]{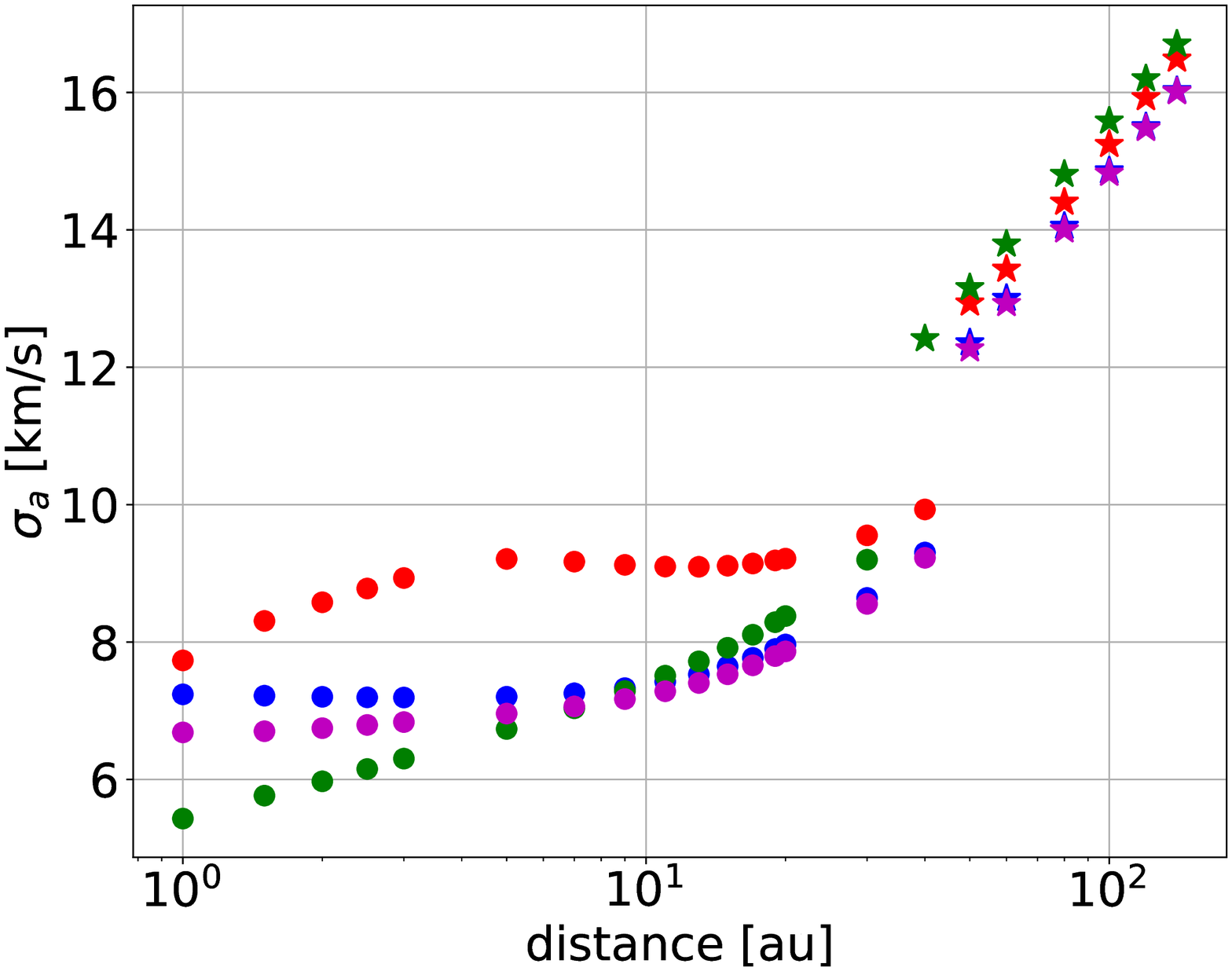}
\caption{The parameters of absorption (the amplitude $A_a$ -- the first row, the mean Doppler shift $\xi_a$ -- the second row, the dispersion $\sigma_a$ -- the third row), presented as a function of heliographic longitude in the heliographic equatorial plane (left column), heliographic latitude in the crosswind plane (center column) and a distance from the Sun for four selected directions (right column). Markers indicate the nodes of our computation grid. Circles correspond to $n_a=2$, while stars to $n_a=4$. A table of all parameters in the 3D grid is available online at: \url{http://users.cbk.waw.pl/~ikowalska/index.php?content=abs}}
\label{fig:params}
\end{figure*}

The amplitude of absorption increases with the distance from the Sun.
It is an expected effect since the column density behaves the same way.
For the same reason, there is also higher amplitude of absorption around the upwind direction.
There is some asymmetry visible between the northern and the southern hemispheres, where the amplitude tends to be higher for northern ecliptic latitudes (see the second plot in the first row in Figure \ref{fig:params}).
The same effect can be seen in the maps presented in Figure \ref{fig:abs_map}.

The magnitude of the Doppler shift of the absorption feature is stable with the distance for a given heliocentric line, as it can be seen in Figure \ref{fig:absorb_prof}.
The modulation of this quantity with the longitude reflects the modulation of the average radial velocity of the gas in different directions.
ISN H flows in from the upwind direction, where the radial velocity of the gas relative to the Sun is negative; hence the shift of the absorption profiles visible in the left panel of Figure \ref{fig:absorb_prof}. At the downwind axis, ISN H flows away from the Sun, and the absorption feature is Doppler-shifted towards positive radial velocities (see top axis in Figure \ref{fig:absorb_prof}). 

The finite width of the absorption profiles, represented by the standard deviation parameter $\sigma_a$ in our fits, exists because of the thermal broadening and a certain radial gradient in the radial component of the velocity of ISN H along this radial line. The broadening is larger for greater distances because the thermal broadening of ISN H is reduced towards the Sun.

In Figure \ref{fig:params}, there are some discontinuity features (see the green markers in the top left panel), which are caused by the fact that we use two alternative fitting functions.
Some of the grid points in this case were fitted by one function (circles), and others by the other one (stars).
The transition occurs when the absorption profiles becomes saturated.
It seems that for the small distances (less than 30 au) the Gaussian function is preferred, while for large distances (greater than 60 au) the function with $n_a=4$ fits better. In the region of intermediate distances we observe a transition between these two functions, which results in an abrupt change in the amplitude parameter.

The simple approximation for the profile of absorption shown in this section can be easily implemented and saves a lot of computational time.
Even far away from the Sun, where saturation is clearly visible, the fitted function corresponds quite well with the numerically simulated profiles.

\clearpage
\subsection{Radiation pressure attenuation factor and optical depth}
\label{sec:attenfact}
In this section, we analyse how absorption affects the gas density and the effective radiation pressure distribution in space.
To that end, we define a parameter that can be used to modify the radiation pressure factor $\mu$, which in the absence of absorption is distance- and location-independent.
By doing so, we will reduce the information that we have about the profile since we eliminate the dependence of radiation pressure on radial velocity of individual H atoms, but facilitate the use of absorption-related modification of radiation pressure in modeling of ISN H distribution in the heliosphere.   

We introduce the effective radiation pressure attenuation factor $f_{abs}(\myvec{r})$ that tells how strong the absorption effect is.
This factor depends on the distance from the Sun and the direction in space described by $\myvec{r}$:
\begin{equation}
\label{eq:f_abs}
   f_{abs}(\myvec{r})=
    \begin{cases}
     \frac{\int\limits_{u_{r_1}}^{u_{r_2}} \mu_{abs}(\myvec{r},u_r') du_r'}{\int\limits_{u_{r_1}}^{u_{r_2}} \mu(\myvec{r}_\text{E},u_r') du_r'}, & \text{ for }\, r > r_\text{E} \\
    1, & \text{ for }\, r \leq r_\text{E}.
    \end{cases}
  \end{equation}

The integration range ($u_{r,1}$, $u_{r,2}$) is defined so that we integrate over the radial velocities for which the absorbed profile $\mu_{abs}(\myvec{r},u_r)$ differs more than $10^{-3}$ from the profile without absorption $\mu(\myvec{r_\text{E}},u_r)$.
In that way, we integrate over that part of the profile that is actually affected by the absorption.
In other words, we integrate over the radial velocity range of the atoms that are within the considered line of sight.
With this definition, the integration region varies from one grid point to another.  

The magnitude of the attenuation factor is limited to 1 (from the definition).
When it is 1, there is no absorption, when it is smaller than 1, absorption is present and thus, there is a reduction in the effective radiation pressure. 

\begin{figure*}[ht!]
\centering
\includegraphics[width=0.6\linewidth]{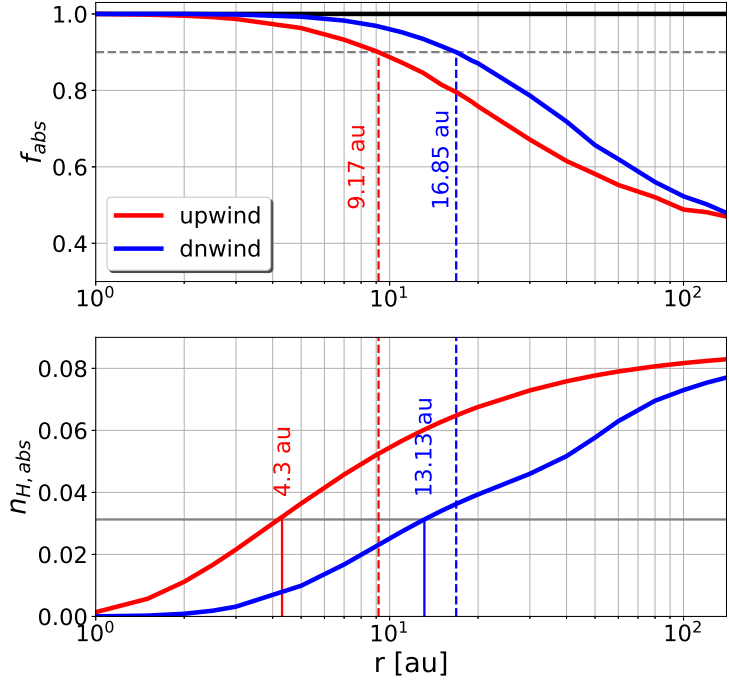}
\caption{Top panel shows the attenuation factor as a function of radial distance, simulated for 1999. Red line shows the upwind direction, blue line shows the downwind direction, and black line corresponds to the case where no absorption is present. The broken vertical bars mark the distances at which the attenuation factor drops to 0.9. Bottom panel presents hydrogen density as a function of radial distance in two directions (upwind - red and downwind - blue). Vertical dashed lines are as on the top panel, while vertical solid lines mark radial distance of the hydrogen cavity (e.g. where local density is $e$ times smaller than on the boundary of calculations). }
\label{fig:f_abs}
\end{figure*}

The flux of electromagnetic radiation in empty space decreases with the distance squared.
But because of the absorption on hydrogen atoms, the solar \lya flux behaves differently.
Figure \ref{fig:f_abs} shows this dependence for two characteristic directions in space: upwind and downwind. The attenuation factor drops to 0.9 at a distance of 9.17 au in the upwind direction and at a distance of 16.85 au in the downwind direction, which is outside hydrogen cavity located at 4.3 au and 13.13 au for upwind and downwind direction, respectively.
There is clear asymmetry in the absorption caused by the asymmetry in the local density distribution.
Close to the Sun, where there is almost no ISN H inside the hydrogen cavity, absorption is almost negligible.
Farther away, the slope becomes steeper as more and more hydrogen atoms  interact with \lya{} photons.

\begin{figure*}[ht!]
\centering
\includegraphics[width=0.45\linewidth]{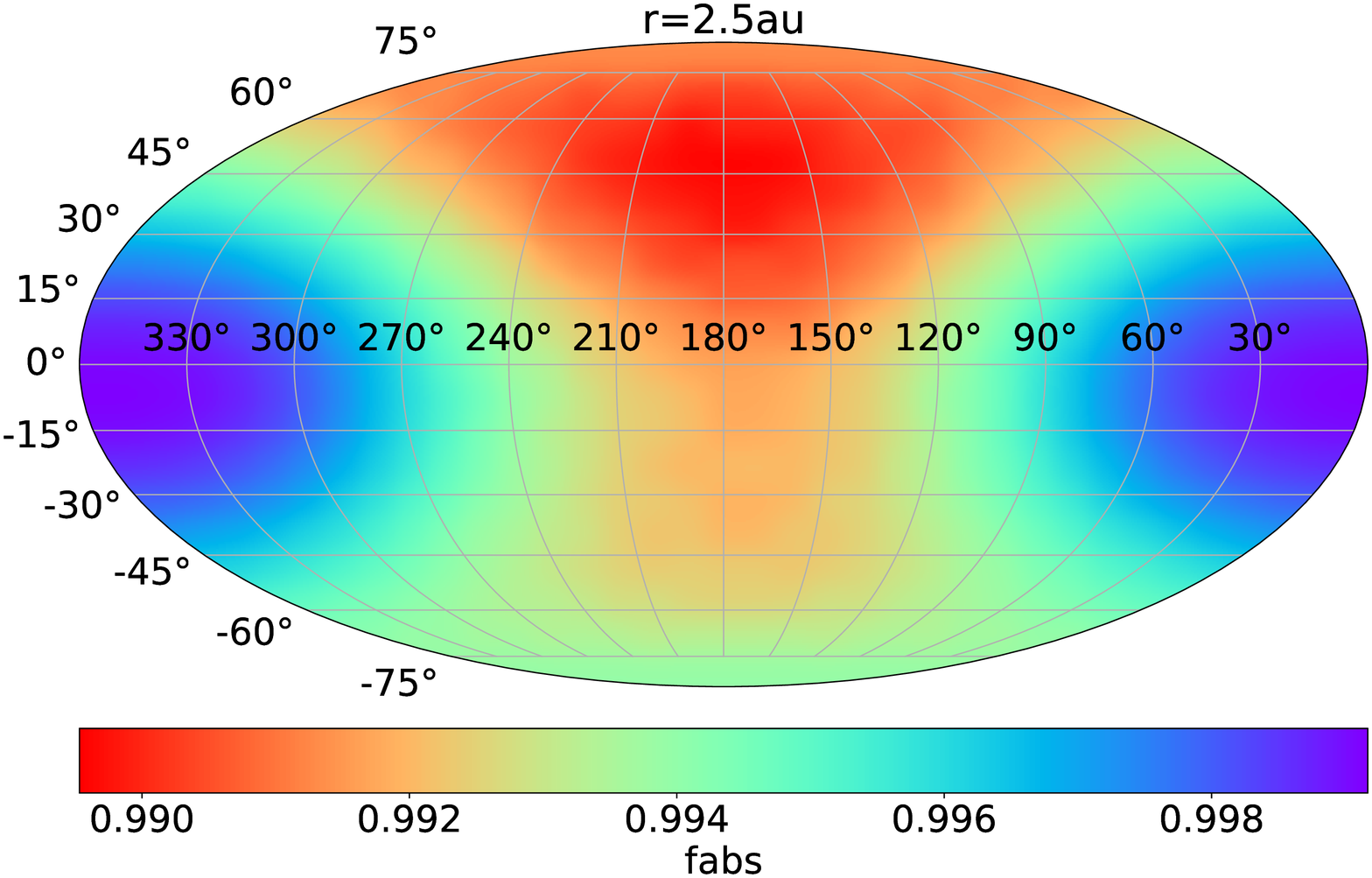}
\includegraphics[width=0.45\linewidth]{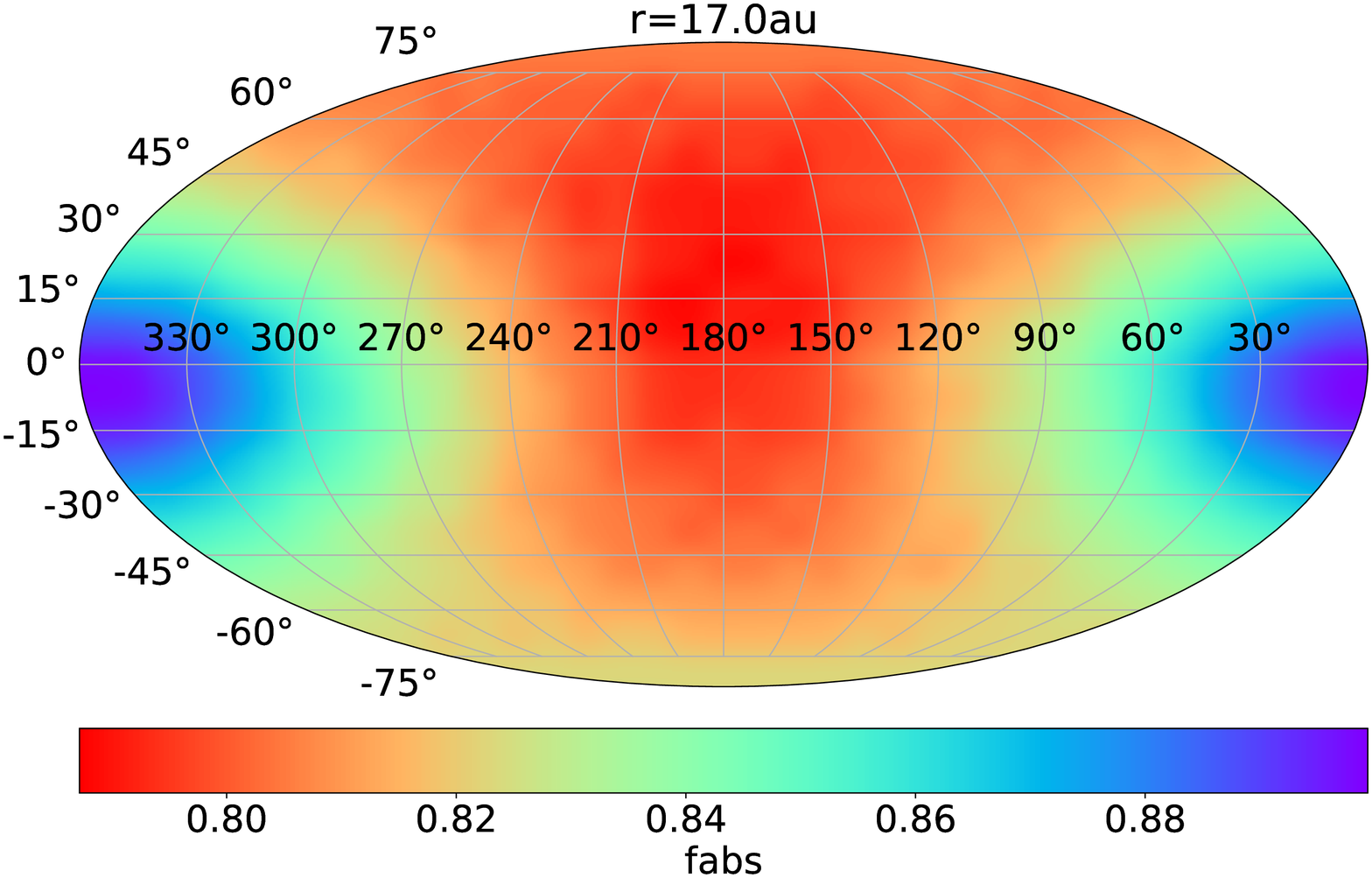}
\includegraphics[width=0.45\linewidth]{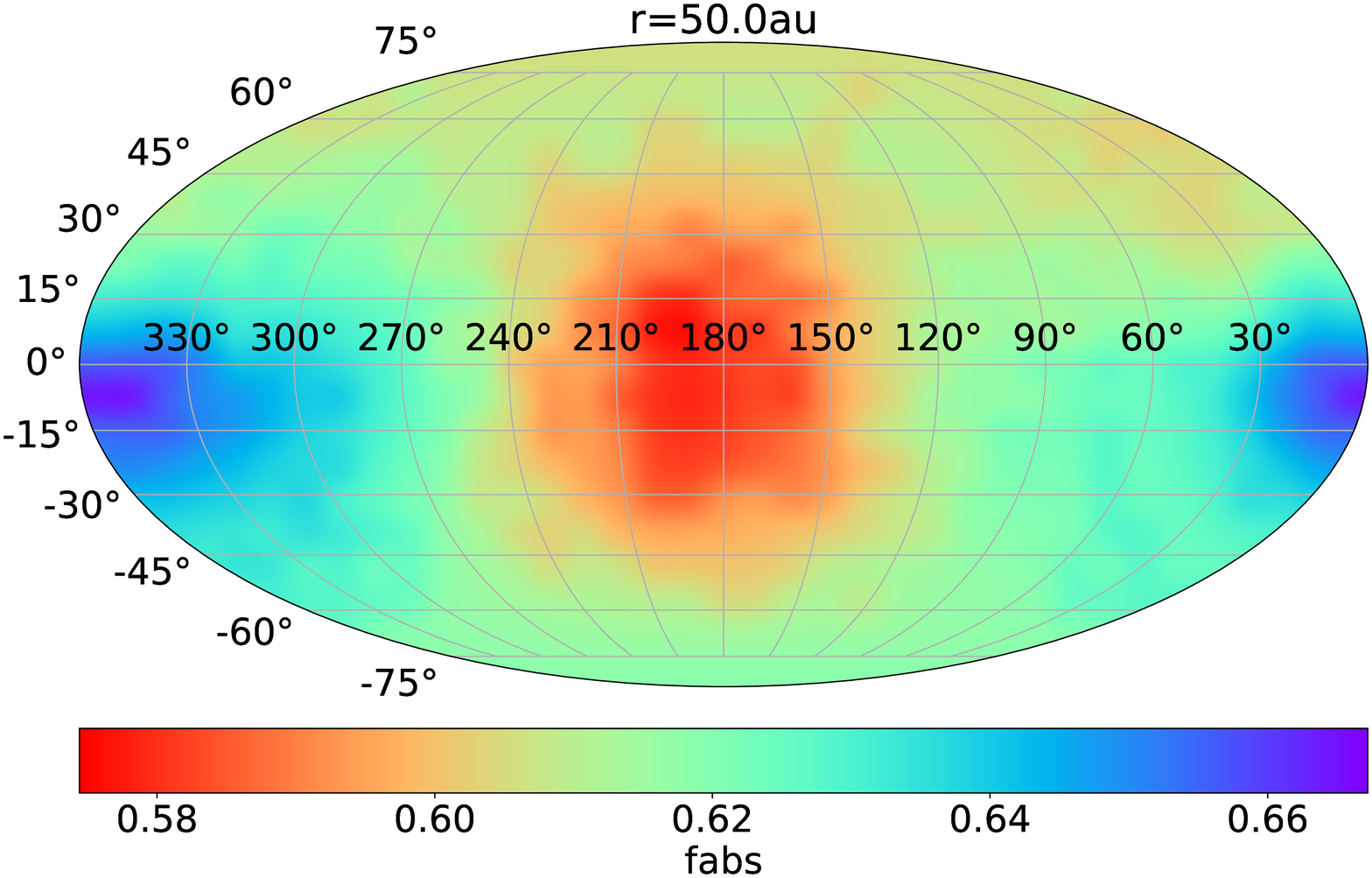}
\includegraphics[width=0.45\linewidth]{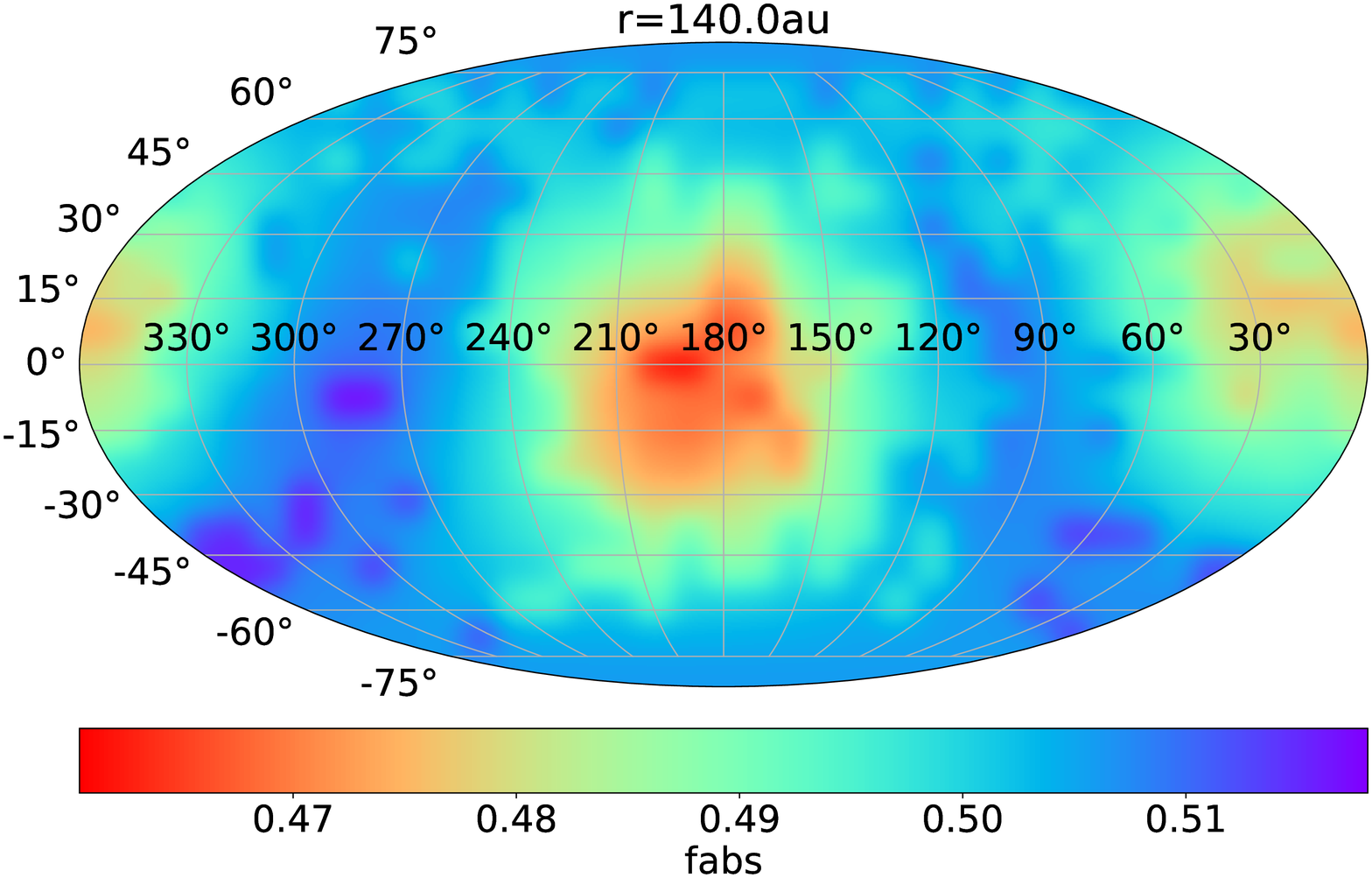}
\caption{Full-sky maps of the attenuation factor at selected distances from the Sun. The maps are shown in the heliographic coordinates. The upwind direction is at $lon=179.346\degr$ and $lat=5.128\degr$}
\label{fig:abs_map}
\end{figure*}

For better visualization, we created 2D maps of the sky at different distances that show the attenuation factor, presented in Figure~\ref{fig:abs_map}.
As expected, the strongest effect is in the upwind direction, where the density of ISN H atoms is the largest (for instance, at 17 au in the upwind direction $f_{abs}=0.79$, and in the downwind direction $f_{abs}=0.89$), even though the strongest effect of absorption on the density is found in the downwind direction (compare Figure \ref{fig:dens_ratio_map}).
We expected this, because the atoms located in the downwind directions have travelled the longest way from the unperturbed interstellar medium through the region where the solar radiation is absorbed. Consequently, they have been subjected to the absorption-reduced radiation pressure for much longer time than the atoms in the upwind direction.
This difference is visible in the density distribution.

Inspection of figure \ref{fig:abs_map} shows that the absorption is the most inhomogeneous at middle distances, represented by the map for 17 au.
In this map, there is a clear structure of absorption related to the heliolatitude structure of the solar wind and resulting heliolatitude structure of the density of ISN H.
The smallest attenuation factor is about 0.78, and the largest about 0.9.
While the magnitude of absorption coefficient is relatively small, its inhomogeneity by percentage is relatively large.
By contrast, at larger distances, the magnitude of absorption is larger, with typical magnitudes of the absorption coefficient 0.62 at 50 au, reaching up to 0.58 in the upwind direction and going down to 0.66 in a relatively narrow region around the downwind axis.
At 140 au, the magnitude of absorption is still larger, with a typical value of the attenuation factor 0.49 and the span between the largest and lowest values 0.46 and 0.52.

The attenuation factor is an approximation for the absorption effect for the solar spectral flux, but can be also used to determine how we need to modify the average radiation pressure at a given point in space. The part of the \lya{} profile that is responsible for the radiation pressure force acting on the considered hydrogen atoms is affected by the absorption effect. To calculate the averaged radiation pressure coefficient $\bar{\mu}_\text{E}(t)$, we need to multiply the averaged radiation pressure measured at 1 au by the attenuation factor. 
\begin{equation}
\label{eq_mu}
\bar{\mu}(\myvec{r},t)=\bar{\mu}_\text{E}(t) f_{abs}(\myvec{r},t),
\end{equation}
where $\myvec{r}$ indicates the position with respect to the Sun in the heliographic coordinate system.
It is important to average radiation pressure at 1 au in the same limits as those used to calculate the corresponding $f_{abs}$. The limits used in our calculations along with the values of the attenuation factor ($f_{abs}$) are available in supplementary table in the MRT standard (see Table \ref{tab:params} as an example) as well as on the web page\footnote{http://users.cbk.waw.pl/~ikowalska/index.php?content=abs}.
In our model \citep{IKL:18a}, the spectral flux at 1 au is axially-symmetric with the symmetry axis being the polar line and it depends only on heliographic latitude. By time dependence we mean that the results depends on the solar cycle phase.

\begin{deluxetable*}{cccccccccc}
\tablecaption{\label{tab:params} Fit parameters and attenuation factor along with integration limits calculated for each nod of our computation grid. Full table is available as a text file. This is only an example.
}
\tablehead{\colhead{R [au]} & \colhead{ELON [deg]} & \colhead{ELAT [deg]} & \colhead{$A_a$} & \colhead{$\xi_a$ [km s$^-1$]} & \colhead{$\sigma_a$ [km s$^-1$]} & \colhead{$n_a$} & \colhead{$f_{abs}$} & \colhead{$u_{r,1}$ [km s$^-1$]} & \colhead{$u_{r,2}$ [km s$^-1$]}}
\startdata
 50.00 &  94.883& -68.777& 1.0234 &  8.7801 & 9.9803& 2& 0.6255 &-24.0 & 40.0\\
 50.00 & 88.090& -59.216& 1.0101 & 11.9127&  9.9974& 2& 0.6298 &-21.0  &43.0\\
 50.00 &  84.313& -49.457& 0.9921 & 14.6745& 10.0252& 2& 0.6338 &-19.0 & 45.0\\
 50.00 & 81.805& -39.617& 0.9680&  16.9631& 10.0553 &2& 0.6404 &-17.0  &47.0\\
 50.00 & 79.927& -29.736& 0.9374 & 18.6714& 10.0874 &2 &0.6503 &-15.0&  49.0\\
 50.00 & 78.390& -19.833& 0.9033&  19.7098& 10.1354 &2& 0.6540& -14.0 & 49.0\\
 50.00 & 77.035&  -9.919& 0.7758 & 19.7819& 12.8085 &4& 0.6667& -14.0 & 50.0\\
 50.00 & 75.760 &  0.000& 0.7779 & 19.5226& 12.9347 &4 &0.6568& -14.0 & 49.0\\
 50.00 & 74.485 &  9.919& 0.8034 & 18.6867& 13.1120& 4 &0.6414& -15.0&  48.0\\
 50.00 & 73.130 & 19.833& 0.8340&  17.1997& 13.2702& 4 &0.6308& -17.0 & 47.0\\
 50.00 & 71.593 & 29.736& 0.8601&  15.0992& 13.3645 &4 &0.6180& -19.0 & 45.0\\
 50.00 & 69.715 & 39.617& 0.8803 & 12.4714& 13.3728 &4 &0.6166& -22.0 & 43.0\\
 50.00 & 67.207 & 49.457 &0.8959 &  9.4237& 13.2939& 4 &0.6125 &-25.0 & 40.0\\
\enddata
\end{deluxetable*}

\begin{figure*}[ht!]
\centering
\includegraphics[width=0.45\linewidth]{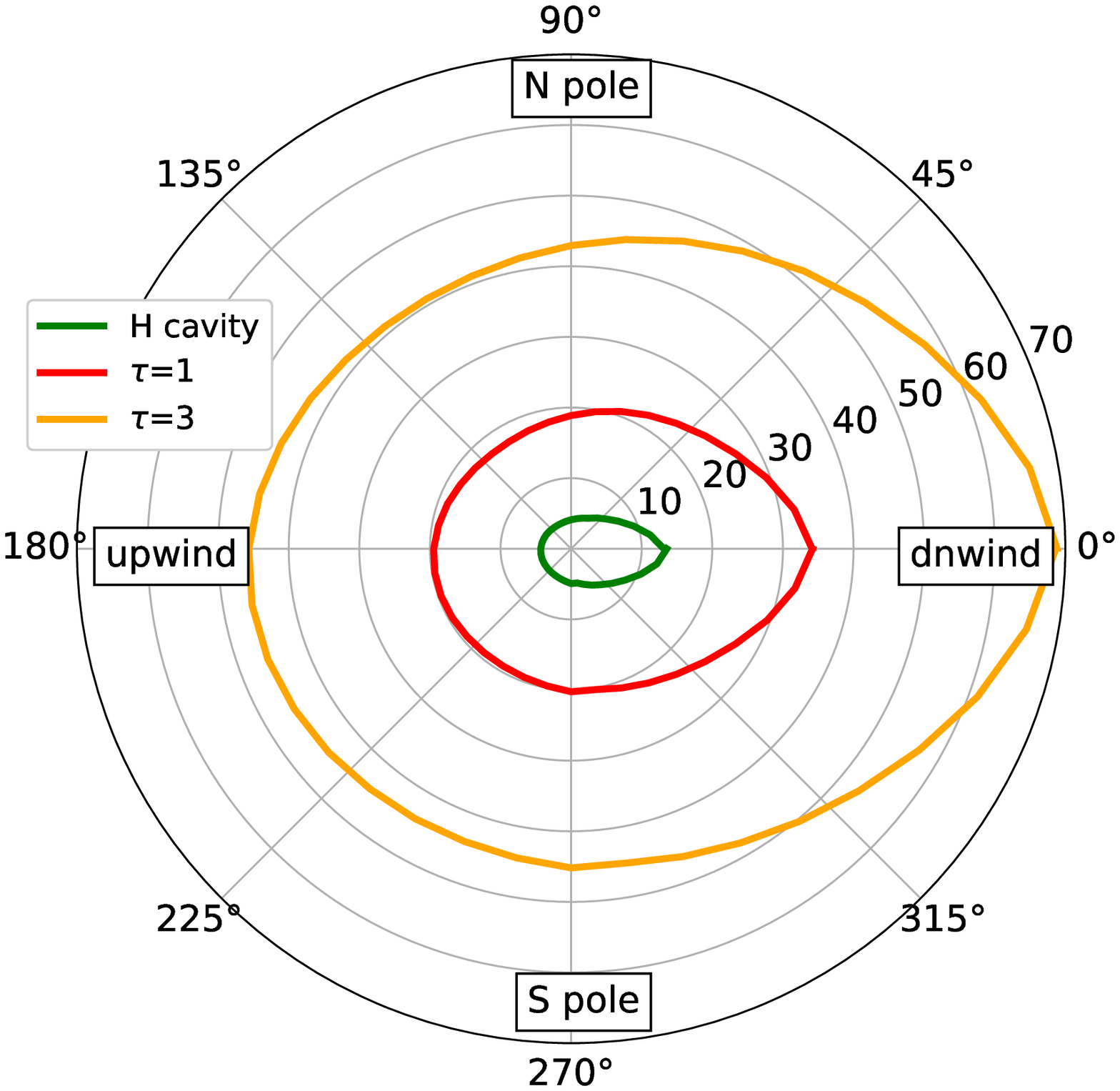}
\includegraphics[width=0.45\linewidth]{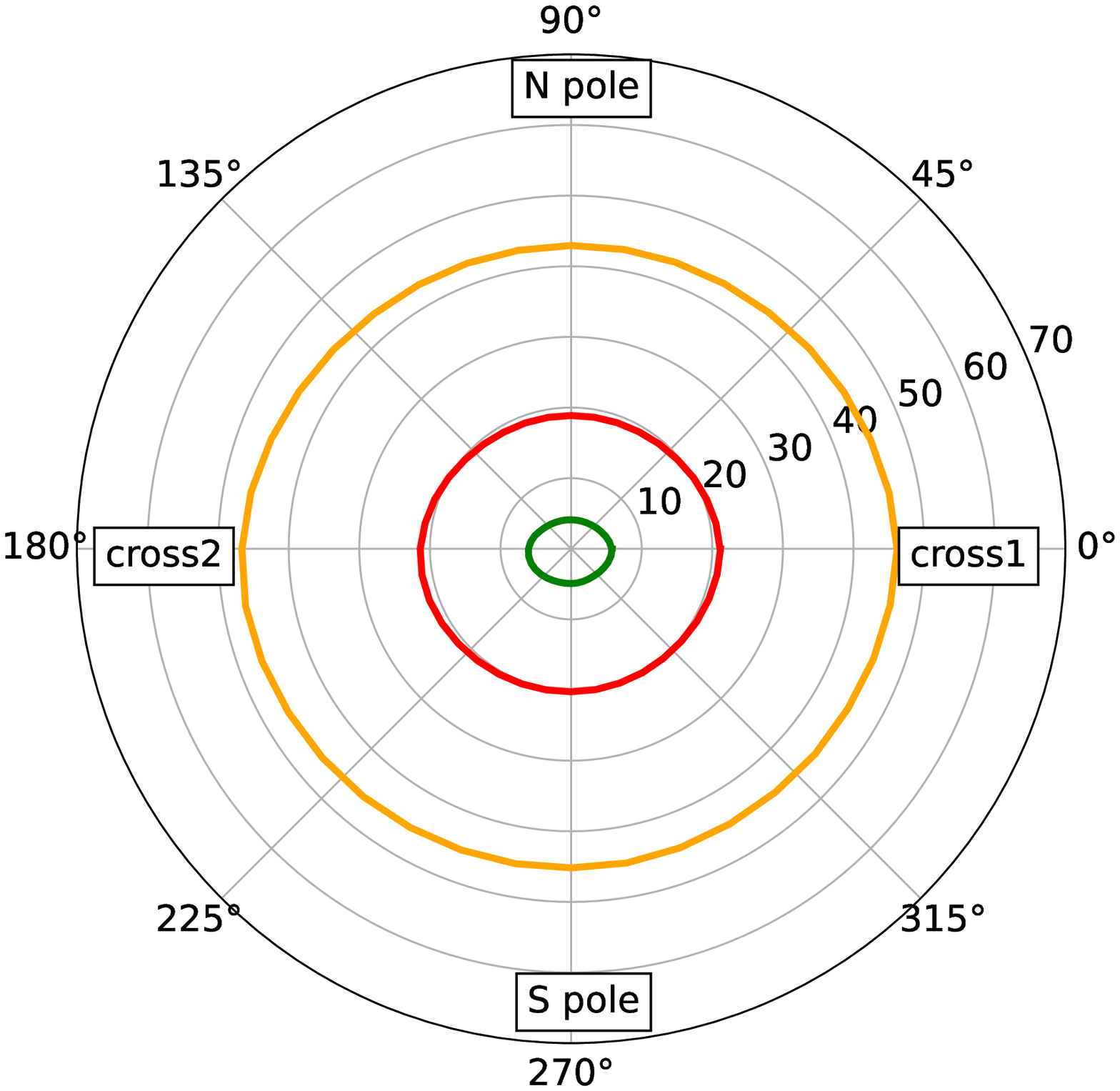}
\includegraphics[width=0.45\linewidth]{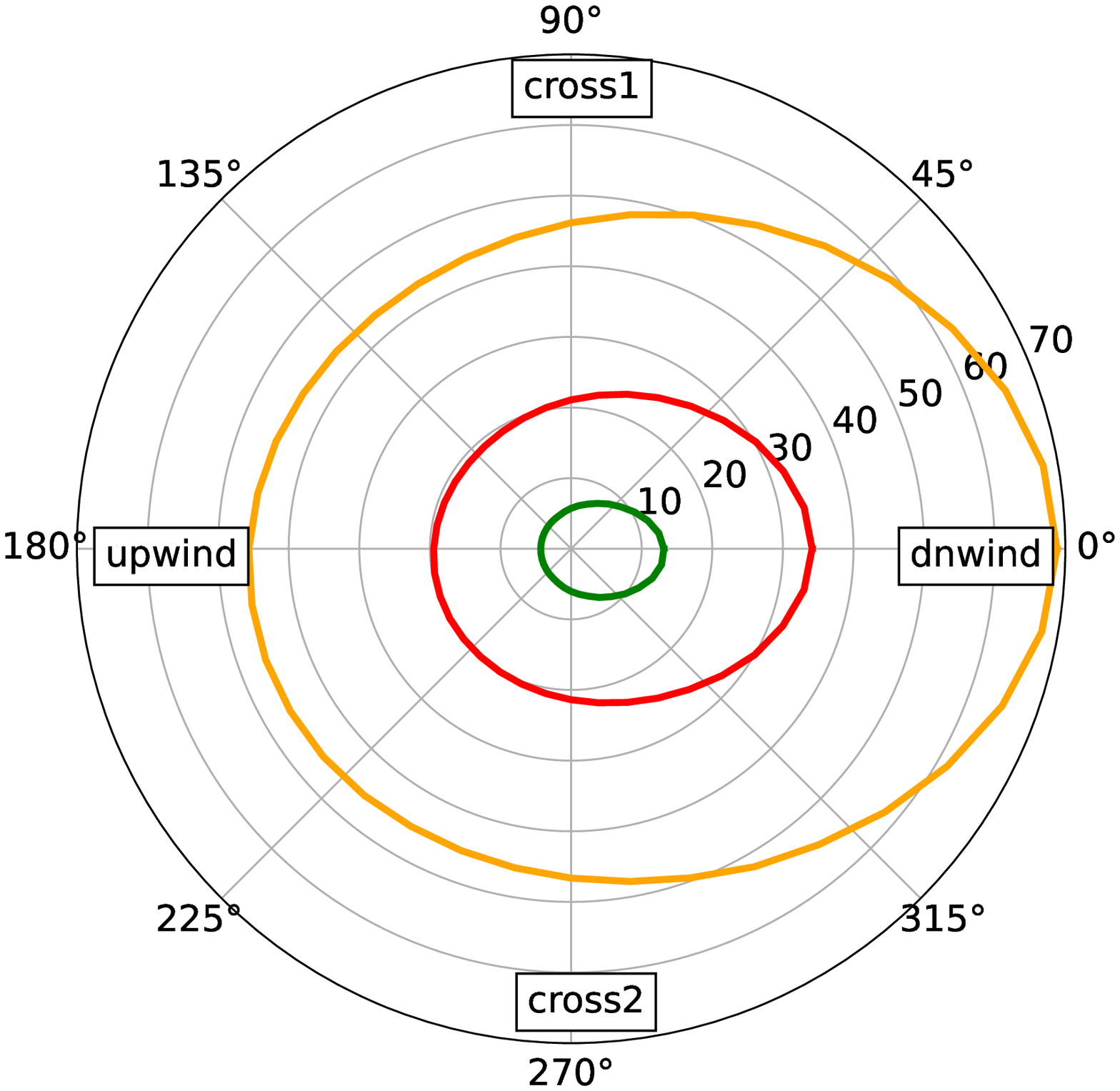}
\caption{Three planes: polar -- the upper left panel, crosswind -- the upper right panel, and equatorial -- the bottom panel. Green line shows the location of the hydrogen cavity, red line shows the location where $\tau=1$, and orange line shows where $\tau=3$. The lines of sight to calculate $\tau$ were assumed radial, originating at 1 au.}
\label{fig:optical_depth}
\end{figure*}

Another parameter commonly used in the literature is the optical depth in the wavelength $\lambda$, which can be expressed as:
\begin{equation}
\label{eg:tau}
\tau_\lambda=\int^r_{0} n_{pr}(\myvec{r'})\sigma_{cs}(\myvec{r'},\Delta \lambda,T_{g,pr}) dr'+\int^r_{0} n_{sc}(\myvec{r'})\sigma_{cs}(\myvec{r'},\Delta \lambda,T_{g,sc}) dr'
\end{equation}
Note that it is calculated for the two-population model, as in Equation \ref{eq:I_abs}.
When the optical depth $\tau=1$, the medium is considered as optical thick. In Figure \ref{fig:optical_depth} we show three different planes with contours corresponding to $\tau=1$ and $\tau=3$ for the wavelengths of maximum absorption. On the same plots we marked the location of the hydrogen cavity. The boundary between optical thin and optical thick medium is shown as a red line. It is located at a distance between 17 and 35 au, depending on the direction. The hydrogen cavity is deep inside the optically thin region, so most of a line of sight for an observer at 1 au traverses the optically thin region. A portion of the signal originating in the optically thick region is relatively low \citep[cf.][]{kubiak_etal:21b}.

\clearpage
\subsection{Hydrogen density}
\label{sec:Hdensity}
The distribution of the ISN H is calculated using the nWTPM \citep{tarnopolski_bzowski:09} with subsequent modifications recapitulated by \citet{IKL:18b} and with the effect of absorption introduced in this paper. 

The radiation pressure acting on hydrogen atoms is calculated using our latest model of the relation between the total flux in the solar \lya line and the spectral flux within this line \citep{IKL:20a}.
We used results from this calculation to modify the radiation pressure that depends on the absorption effect.
It was done iteratively (as it was described in Section \ref{sec:calculations}).

\begin{figure*}[ht!]
\centering
\includegraphics[width=0.45\linewidth]{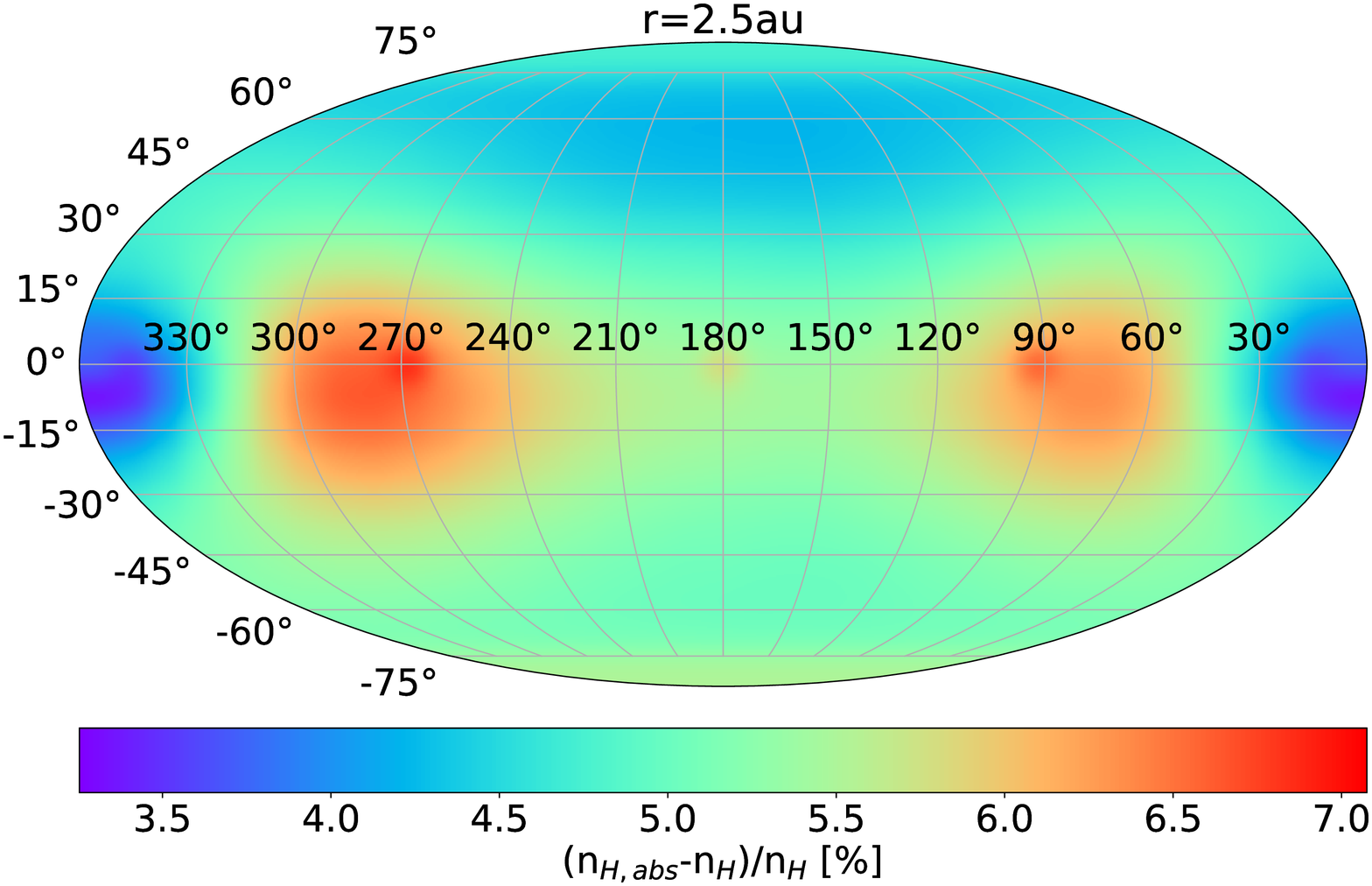}
\includegraphics[width=0.45\linewidth]{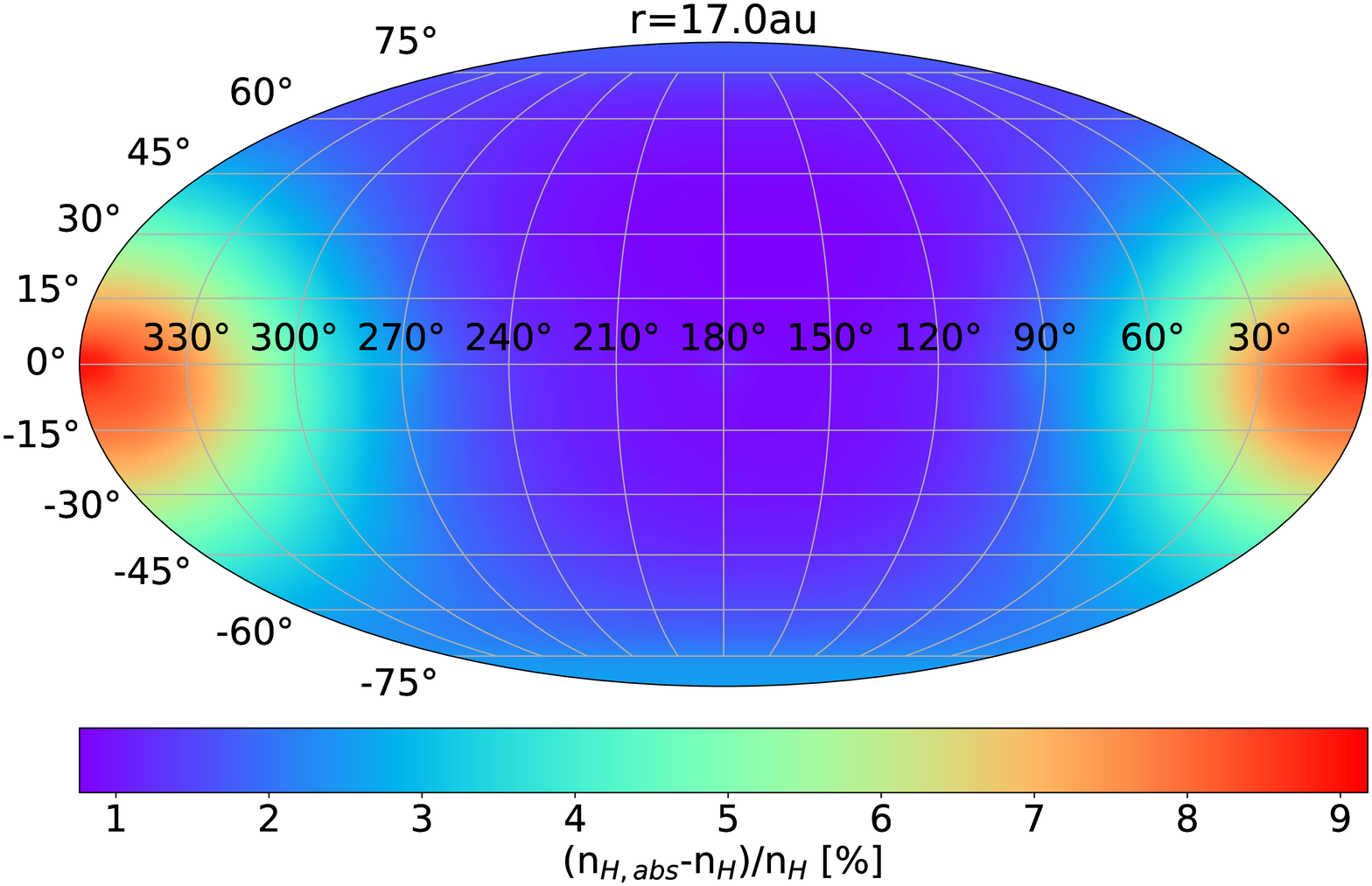}
\includegraphics[width=0.45\linewidth]{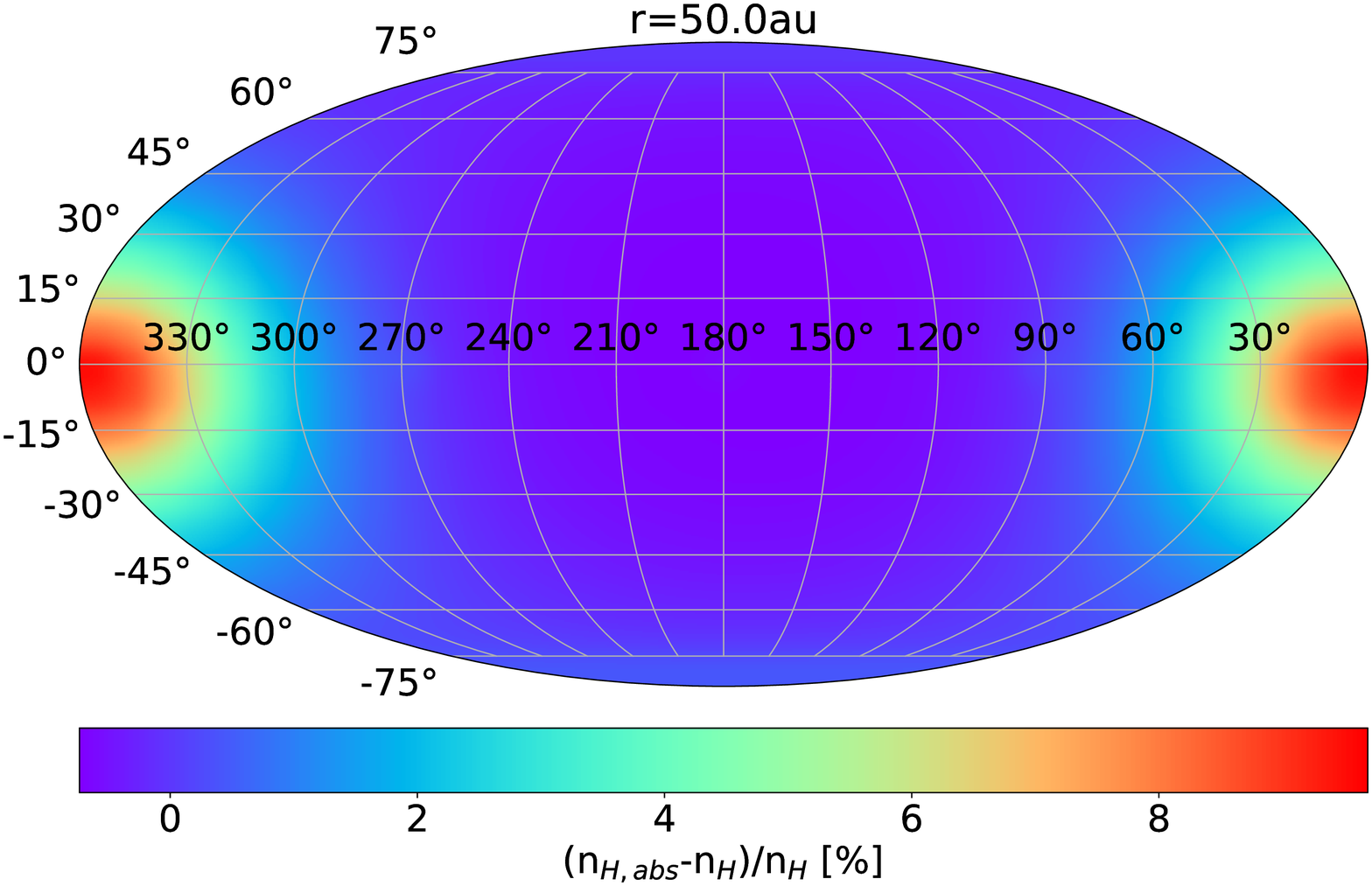}
\includegraphics[width=0.45\linewidth]{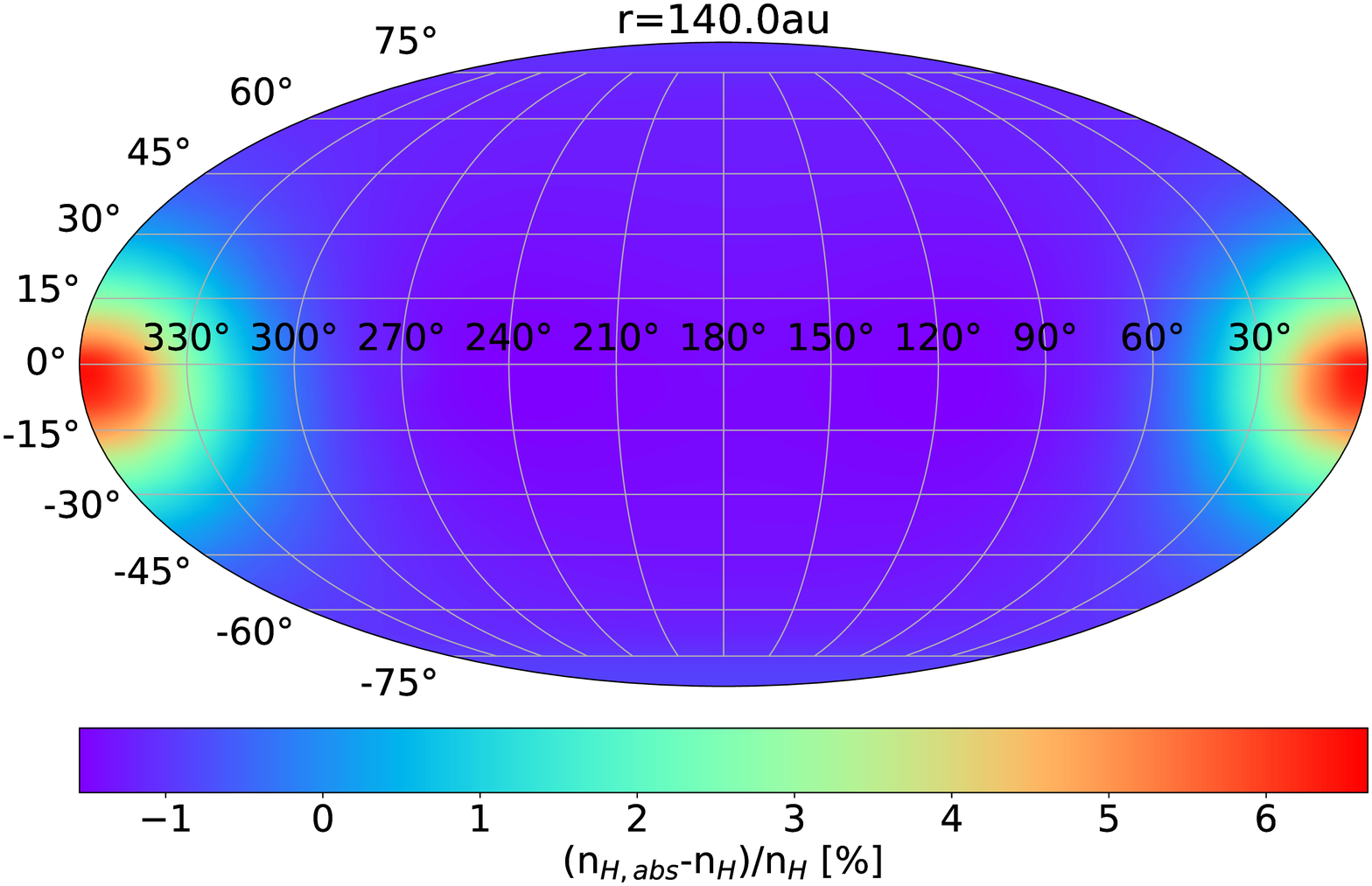}
\caption{Maps of the difference between density calculated with absorption included and density without absorption normalized by the density without absorption. The coordinate system is the same as in Figure \ref{fig:abs_map}.}
\label{fig:dens_ratio_map}
\end{figure*}

We show the difference between the density calculated with the absorption effect included ($n_{\text{H},abs}$) and the density without this effect ($n_\text{H}$).
It will be helpful in identifying the regions in the heliosphere where absorption has a big impact and should not be neglected.
Figure \ref{fig:dens_ratio_map} demonstrates that neglecting the absorption results in underestimating of the density by $\sim 9$\% in the downwind direction, i.e., in the worst case.
A strong effect is also visible near the Sun, but the absolute values of the density in this region are very small.
It is not without significance that the map for the distance of 2.5 au is located inside the hydrogen cavity.

\clearpage
\subsection{Dependence on the phase of the solar cycle}
\label{sec:solar_cycle}
\begin{figure*}[ht!]
\centering
\includegraphics[width=0.95\linewidth]{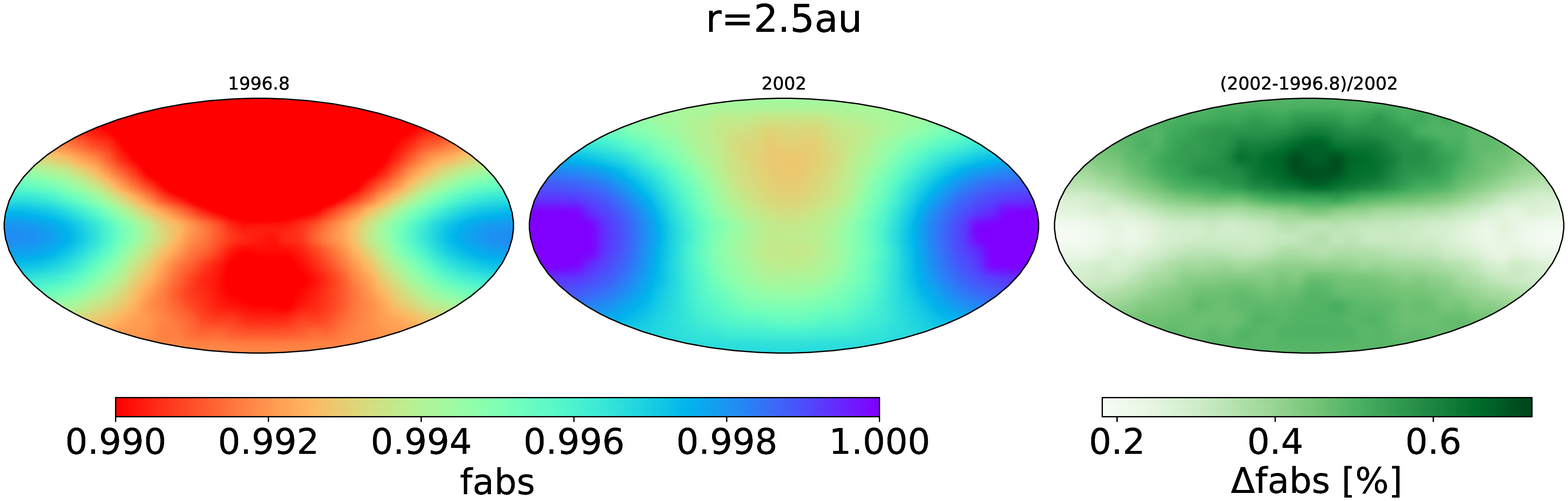}
\includegraphics[width=0.95\linewidth]{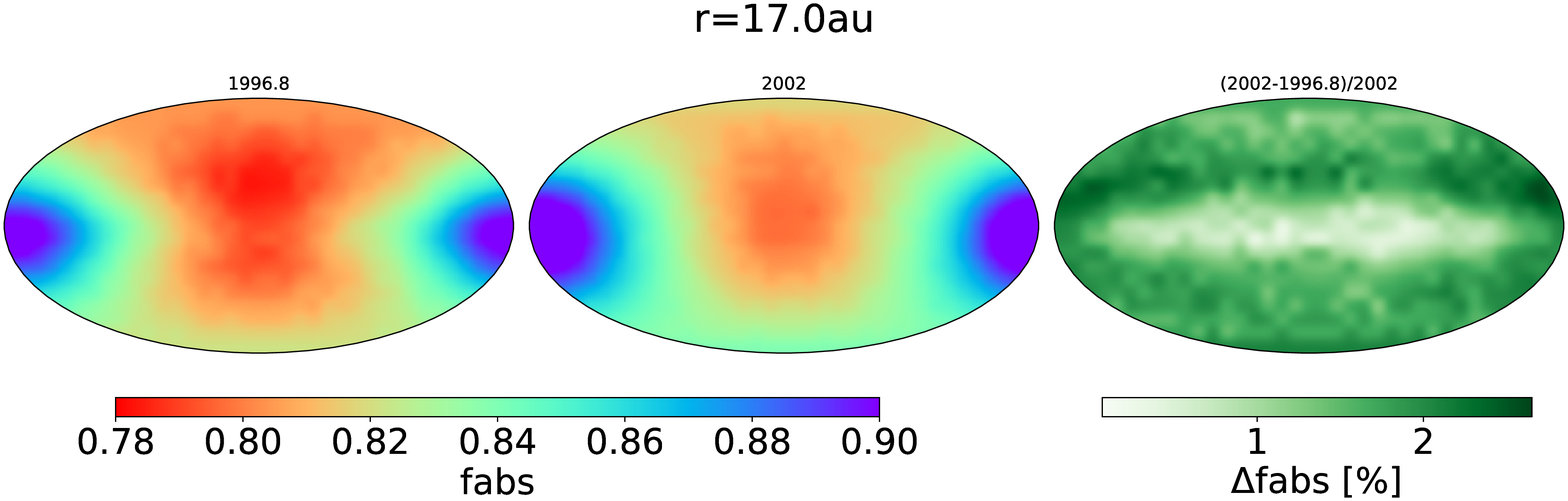}
\caption{Full-sky maps of the attenuation factor at selected distances from the Sun for the epochs 1996.8 (left column) and 2002.0 (the middle column). The third column shows the percentage difference between these two maps. Top row shows a distance of 2.5 au, while the bottom row shows this for 17 au. The maps are shown in heliographic coordinates.}
\label{fig:minMaxMap1}
\end{figure*}

\begin{figure*}[ht!]
\centering
\includegraphics[width=0.95\linewidth]{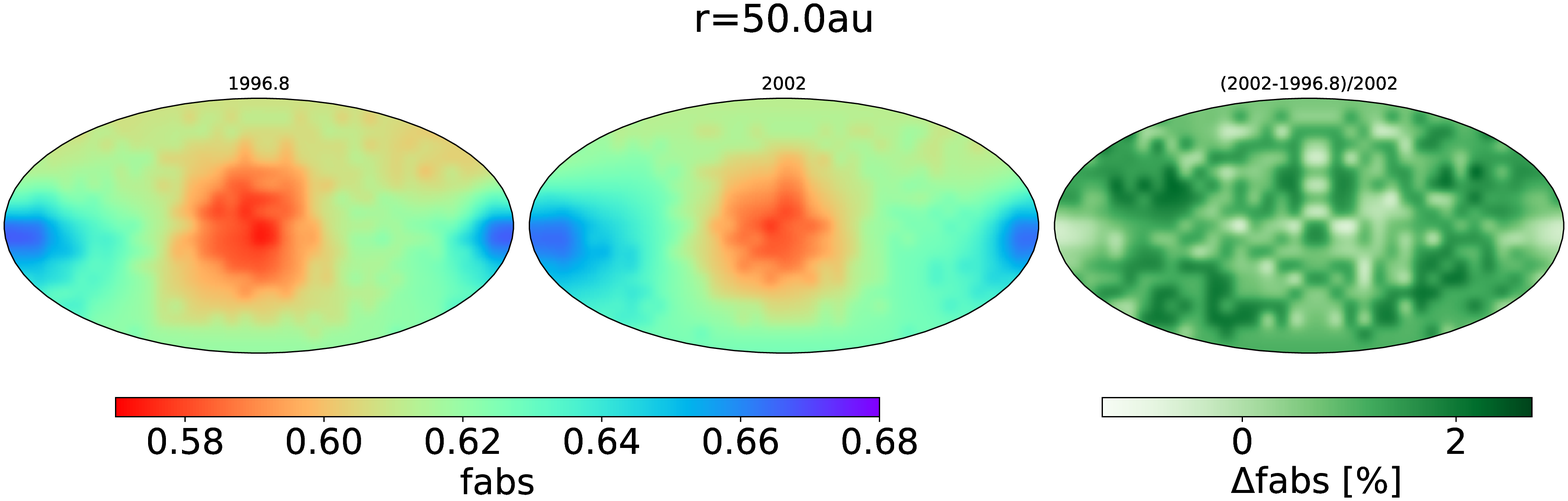}
\includegraphics[width=0.95\linewidth]{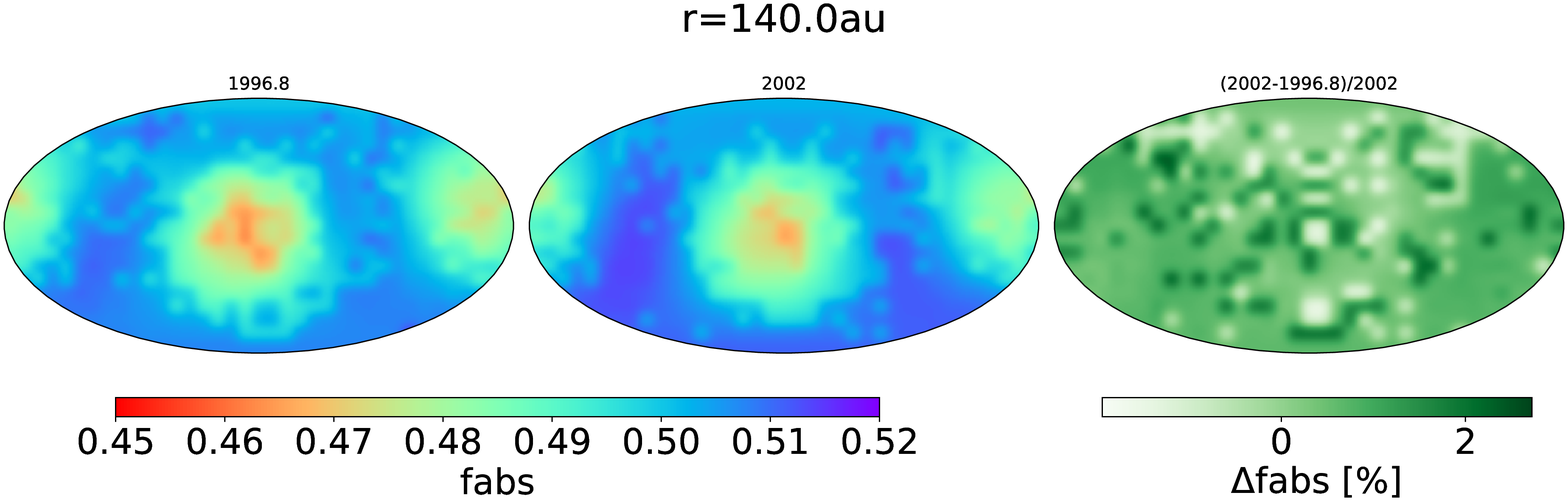}
\caption{Same as Figure \ref{fig:minMaxMap1}, but for distance 50 au (top row) and 140 au (bottom row)}
\label{fig:minMaxMap2}
\end{figure*}

In the previous section, we showed simulations performed for one moment in time (1999.0, when the solar activity was on an average level) to illustrate how the absorption effect works for different directions and distances in the heliosphere.
However, ISN H atoms need a substantial time to travel through the heliosphere, longer than the length of the solar cycle \citep{bzowski_kubiak:20a}.
During this travel, the radiation pressure that acts on H atoms is caused by the Sun in different phases of its activity.
Hence,  we need an estimate of absorption for various phases of the solar cycle.
The effect of absorption depends on the intensity of the \lya radiation and the column density of the ISN H.
The first one is modulated by the solar cycle activity, the second one is much more complex and in addition to radiation pressure, it also depends on the solar wind conditions and the photoionization rate.
The time evolution of these factors is only partly periodic and there is no simple model that allows us to make an analytical or even semi-analytical parametrization.
We compared the absorption effects in two extreme cases, when the local density reaches its maximum and minimum values during the solar cycle.
The local density at 1 au in the upwind direction has the highest value close to the solar minimum conditions (in our simulations this minimum was in November 1996) and has the lowest value during the solar maximum (in January 2002).
Figures \ref{fig:minMaxMap1} and \ref{fig:minMaxMap2} show the difference between the magnitudes of the attenuation factor in these two extreme cases. 

Inside the hydrogen cavity (the first row of plots in Figure \ref{fig:minMaxMap1}), the absorption effect is very weak and it is changing during solar cycle by a fraction of percent.
Further away from the Sun, the amplitude of changes during the solar activity cycle is larger, but never exceeds 3\%.
Therefore, all results presented in this paper (calculated for the mean solar activity in 1999) are accurate within 3\% uncertainty.
\clearpage
\subsection{Dependence of absorption on the density beyond the termination shock}
\label{sec:TSdens}
A recent analysis \citep{swaczyna_etal:20a} has shown that density of the hydrogen on the TS is not known as well as we thought \citep{bzowski_etal:09a}.
The difference in the estimates is 49\%, i.e., quite substantial.
We analyzed the effect of changing the density at the boundary of our calculations on the magnitude of absorption. 

\begin{figure*}[ht!]
\centering
\includegraphics[width=0.6\linewidth]{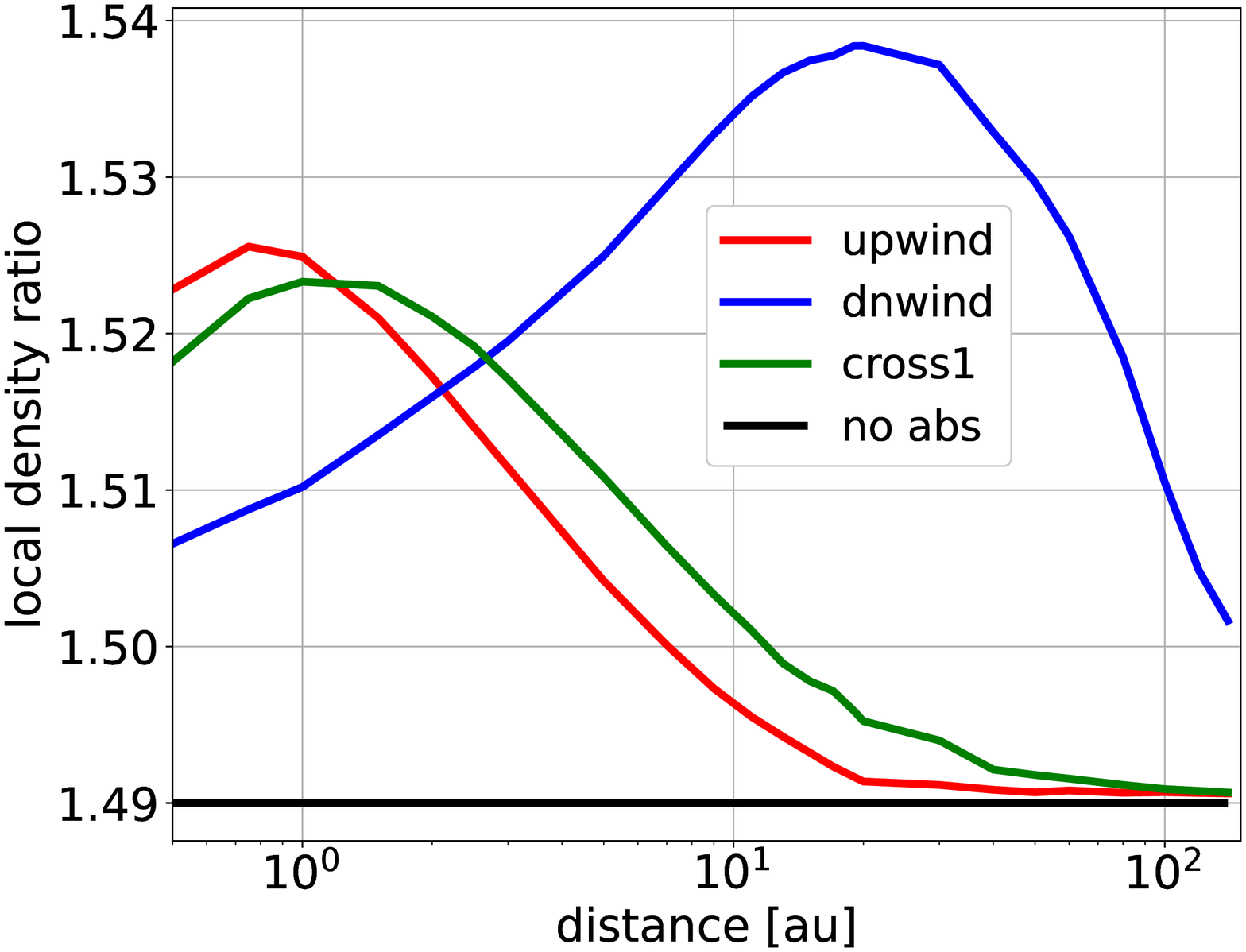}
\caption{Ratio of local density with different initial density at the boundary of our calculations. The higher value is 0.127 and the smaller one is 0.085. Absorption is included. Colours show different directions: red line -- upwind, blue line -- downwind, green line -- crosswind. Black horizontal line shows the constant ratio when we considering no-absorption case.}
\label{fig:TSdens}
\end{figure*}

In our simulations of the gas distribution, the density at 300 au (the boundary of the calculations) scales the density inside the heliosphere linearly.
If we ignore absorption, the ratio of the local densities calculated for different densities at 300 au is constant (the black line in Figure \ref{fig:TSdens}).
But when we include absorption in our simulations, the ratio of the local density is changing with the distance from the Sun, and the gradient of these changes depends on the direction.
For an upwind direction, the ratio is higher than in the absorption-free case near the Sun and it is decreasing with the distance to stabilize around 20 au (the red line in Figure \ref{fig:TSdens}).
Similar situation is in crosswind direction (green line in Figure \ref{fig:TSdens}), but stabilization is achieved much further from the Sun (around 40 au).
A different behavior is seen for the downwind direction (the blue line in Figure \ref{fig:TSdens}), where the density ratio increases for the first 10 au and then slowly decreases with the distance, but never achieves the stabilization level characteristic for absorption-free case.
This can be explained by general low density in the tail, but also by the fact that in this region we expect not only a direct flow of the ISN H, but also inverse beam.

The absorption effect is stronger when the assumed density at the TS is higher, because the local density is higher.
Adoption of a greater density at the boundary, in conjunction with absorption, results in a small global increase in the densities inside the heliosphere, varying in space.
However, for the range of TS densities in question (0.085 cm$^{-3}$ in \citet{bzowski_etal:09a} vs 0.127 cm$^{-3}$ in \citet{swaczyna_etal:20a}) this increase of the local density is at most by 3\%, and typically smaller. 

\clearpage
\subsection{The signal observed by IBEX-Lo}
\label{sec:IBEXSignal}
In this section, we analyse the quantities relevant for direct-sampling observations of ISN H by IBEX-Lo. IBEX \citep{mccomas_etal:09a} is the first mission to directly sample ISN H in the Earth's orbit.
IBEX is a spin-stabilized spacecraft with the spin axis changed once (until 2012) or twice per orbit to approximately follow the Sun \citep{mccomas_etal:11a}.
The ISN atoms are observed using the IBEX-Lo instrument \citep{fuselier_etal:09b}, which is a time of flight mass spectrometer.
Before entering the detector, the observed atoms pass the collimator, which defines the field of view of the instrument.
The data are collected while the spacecraft is rotating, and the observed counts are binned into time-intervals with the length selected so that they correspond to fixed spin angle bins.
Since between the repositioning the spin axis is fixed in space, the observations during an individual orbital arc cover a specific, fixed region in the sky.

The signal observed by IBEX-Lo depends on the local density and speed of ISN H, which depend on the radiation pressure, so it may be a good tool to constrain our models.
In the past, numerous authors  reported an important  discrepancy between theoretical simulations and the actual IBEX observations of the ISN H.
They suggested that the reason may be that radiation pressure models are not accurate enough (\citet{schwadron_etal:13a,katushkina_etal:15b, rahmanifard_etal:19a}).
We developed a new solar \lya radiation pressure  model to address that issue \citep{IKL:18a}.
In this study we show yet another factor that may help to solve this problem.

In the simulations carried out using the nWTPM code \citep{sokol_etal:15b}, the distribution function of ISN H atoms is represented by a superposition of distribution functions corresponding to the primary and secondary populations of ISN H, and the signal observed by IBEX is simulated for the time of detection and the location and velocity of the detector in space.
In the presented simulations, we show the total flux of ISN H entering the detector, the mean speeds and energies of the two populations separately for spin angle bins identical to these used by IBEX-Lo for IBEX orbits belonging to a selected ISN observation season, which begins in December and ends in April each year.

Even though the ISN observations are carried out during yearly seasons, the observed signal features large variations between the seasons.
The ISN H is heavily modulated in the Earth's orbit due to the variations in the ionization rate and radiation pressure during the solar activity cycle \citep[][]{rucinski_bzowski:95b, bzowski_etal:97, tarnopolski_bzowski:09, bzowski_etal:13a}.
On the other hand, the observation conditions never repeat precisely from one year to another, therefore the observed ISN H signal is very sensitive to small differences in the spin angle pointing and the length of ``good times'' during the observations.

\begin{figure*}
\centering
\includegraphics[width=1.2\textwidth,angle=90]{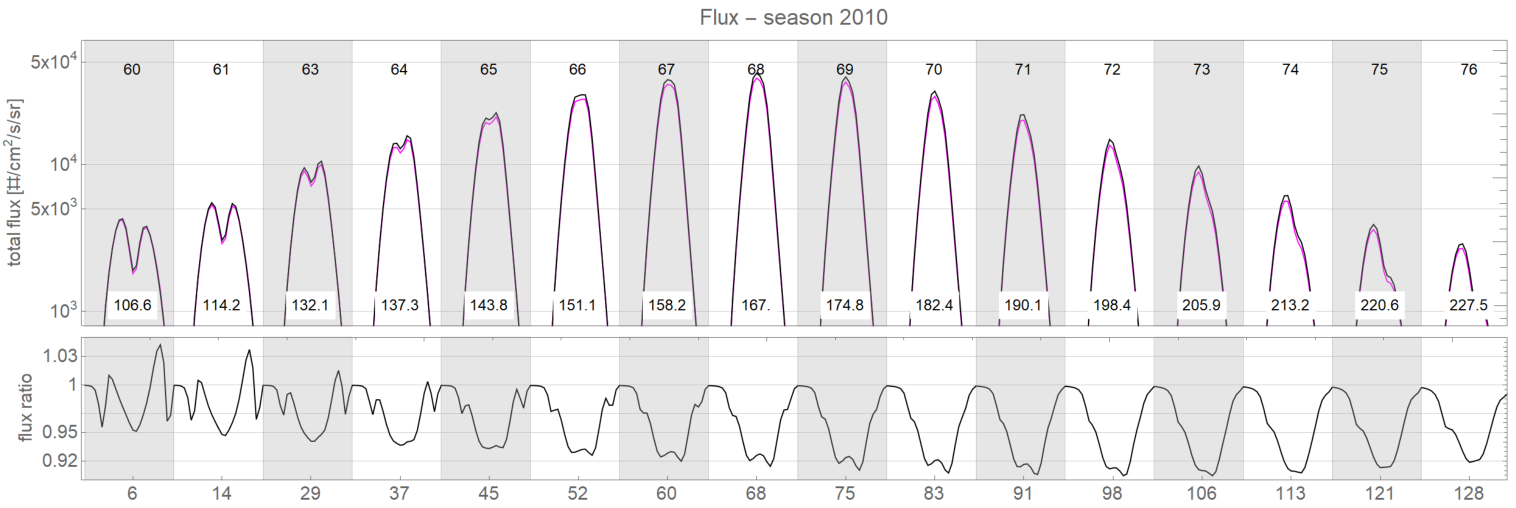}
\includegraphics[width=1.2\textwidth,angle=90]{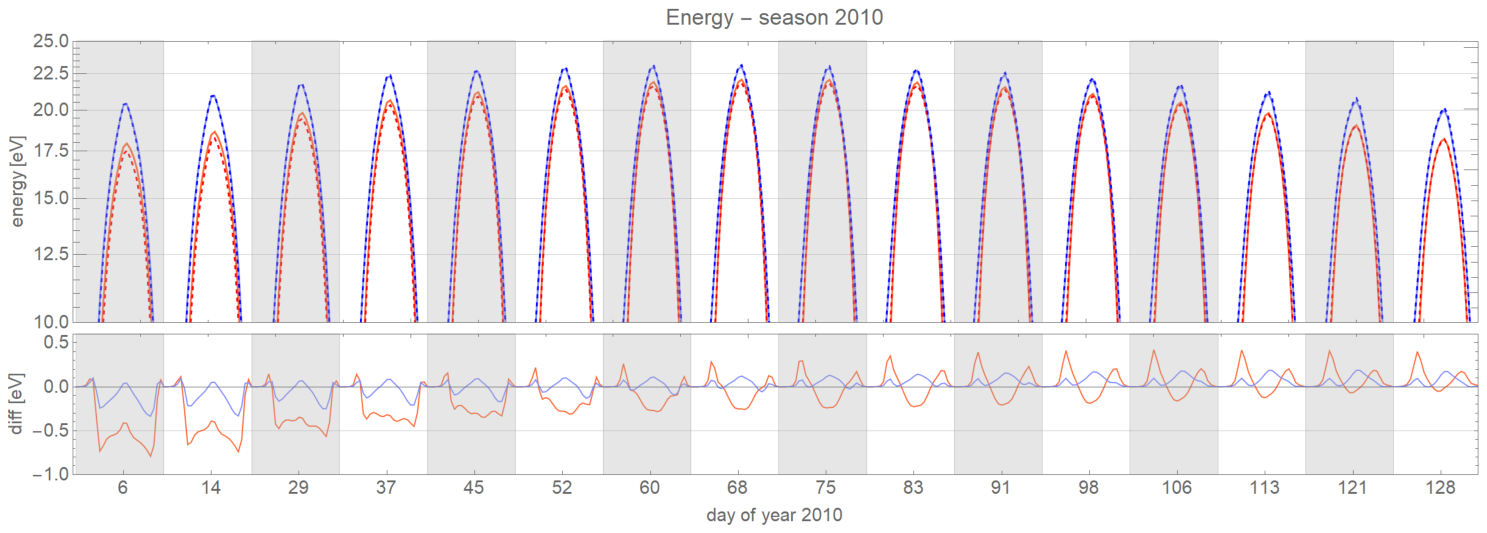}
\caption{
A simulated total flux (a sum of the fluxes of the primary and secondary populations, upper panel), mean energy (separately for the primary and the secondary populations, lower panel) of ISN H filtered by the collimator of IBEX-Lo. The quantities shown were simulated for each 6-degree spin angle interval within the range $180\degr-330\degr$. The black solid line in the upper panel corresponds to the absorbed model and the magenta solid line presents the simulations without absorption. In the lower panels, the red color represents the primary population and blue color the secondary population. Solid lines correspond to the model with absorption, and dashed lines without absorption. Separation between individual orbits is marked by the white and gray strips. The orbit numbers are shown at the top of the first panel, and the mean ecliptic longitudes of the Earth for individual orbits are presented at the bottom of the first panel. The lower subpanel in the first panel represents the ratio of the fluxes without absorption to that with absorption, and the lower subpanel in the second panel shows the differences in the energy, obtained from models with and without absorption.
}
\label{fig:ibex2010}
\end{figure*}

In Figure \ref{fig:ibex2010}, we choose to show the year 2010, which is during the minimum of the solar activity, when the hydrogen flux is the highest.
The biggest absorption effect is seen in the peaks of the middle and latest orbits (68--74), when it reaches up to 9\%.
Usually, the flux simulated with absorption included is higher than that without it.
The magnitude of the effect of absorption on the signal is dependent on the spin angle of the observed bins.
On the wings, it is just around 1\%-4\% but close to the peak flux, where the observation statistics is the best, it is up to 9\%. Since this effect is systematic, it may affect details of the analysis results but is not likely to explain the discrepancy between the model and observation mentioned before. 

Absorption affects the energy of both populations in a similar way, and the magnitude of the effect does not depend on the spin angle or the strength of the signal.

The first part of the season (until orbit 64) gives different absorption-related features than the rest of the signal, namely the flux ratio is sometimes higher than one, and the energy difference is also much lower.
In this region, the signal is built up mostly by the secondary population of ISN H.
This result implies the high sensitivity of the signal observed by IBEX to details of radiation pressure and a large diagnostic potential of this kind of observations.  
\clearpage
\section{Summary and Conclusions}

\lya{} radiation is (next to solar gravity) one of the most important factors influencing the distribution of ISN hydrogen in the heliosphere.
Simulations of radiation pressure in the whole heliosphere are very important to calculate trajectories of hydrogen atoms.
Absorption of the \lya{} radiation modifies the effective radiation pressure profile and the global hydrogen distribution.
In this paper we have calculated the radiation pressure profiles with absorption effects included on a 3D computation grid up to 140 au from the Sun.
The magnitude of absorption that modifies the solar \lya{} profile can be estimated using a Gauss-like function, described by 4 parameters ($A_a$, $\xi_a$, $\sigma_a$ and $n_a$).
We have defined the attenuation factor ($f_{abs}$) that estimates the magnitude of the absorption effect at a given point in the heliosphere.
The attenuation factor increases with an increasing distance from the Sun, thus the \lya{} spectral flux responsible for radiation pressure acting on ISN H atoms in the heliosphere drops faster than $1/r^2$.
We have shown how absorption changes the density of the ISN H. Even though attenuation factor indicates the highest absorption in the upwind direction, relative changes in the hydrogen density are the biggest in around the tail.
In general, absorption modifies the density by several percent, mostly in the downwind cone of $\sim 30\degr$ and at relatively small heliocentric distances of a few au. 
We have analysed modulation of the absorption during solar cycle and found that the changes between extreme conditions were up to 3\%. Therefore, we concluded that our basic set of calculations done for epoch 1999.0 can be widely used within a 3\% accuracy.

Given the uncertainties related to uncertainties in the ionization rate of ISN H, of the solar EUV output etc., absorption may be neglected to the first approximation in the calculation of ISN the H density and related production rates of the pickup ions.
However, we found that absorption systematically modifies the ISN H observed by direct-sampling missions at 1 au, like IBEX and IMAP.
For these observation conditions, systematic differences may be on the order of 5--8\%, and are likely to be larger than the sampling uncertainties.
Therefore, absorption needs to be taken into account in the analyses of these observations, like that presented by \citet{schwadron_etal:13a, katushkina_etal:15b, galli_etal:19a, rahmanifard_etal:19a}.

The absorption effect calculated and described in this paper is incorporated into the latest version of the nWTPM code.

\acknowledgments
{\emph{Acknowledgments}}. 
The authors would like to kindly thank Jeffrey Linsky and Eberhard Moebius for the helpful discussion.
This study was supported by Polish National Science Center grants 2019/35/B/ST9/01241, 2018/31/D/ST9/02852, and by Polish Ministry for Education and Science under contract MEiN/2021/2/DIR.



\bibliographystyle{aasjournal}
\bibliography{iplbib}

\begin{thebibliography}{}
\expandafter\ifx\csname natexlab\endcsname\relax\def\natexlab#1{#1}\fi
\providecommand{\url}[1]{\href{#1}{#1}}
\providecommand{\dodoi}[1]{doi:~\href{http://doi.org/#1}{\nolinkurl{#1}}}
\providecommand{\doeprint}[1]{\href{http://ascl.net/#1}{\nolinkurl{http://ascl.net/#1}}}
\providecommand{\doarXiv}[1]{\href{https://arxiv.org/abs/#1}{\nolinkurl{https://arxiv.org/abs/#1}}}

\bibitem[{Axford(1972)}]{axford:72}
Axford, W.~I. 1972, in The Solar Wind, ed. J.~M.~W. C.~P.~Sonnet, P.
  J.~Coleman, NASA Spec. Publ. 308, 609--660

\bibitem[{Bzowski {et~al.}(1997)Bzowski, Fahr, Ruci{\'n}ski, \&
  Scherer}]{bzowski_etal:97}
Bzowski, M., Fahr, H.~J., Ruci{\'n}ski, D., \& Scherer, H. 1997, \aap, 326, 396

\bibitem[{Bzowski \& Kubiak(2020)}]{bzowski_kubiak:20a}
Bzowski, M., \& Kubiak, M.~A. 2020, \apj, 901, \dodoi{10.3847/1538-4357/abada2}

\bibitem[{{Bzowski} {et~al.}(2009){Bzowski}, {M{\"o}bius}, {Tarnopolski},
  {Izmodenov}, \& {Gloeckler}}]{bzowski_etal:09a}
{Bzowski}, M., {M{\"o}bius}, E., {Tarnopolski}, S., {Izmodenov}, V., \&
  {Gloeckler}, G. 2009, \ssr, 143, 177, \dodoi{10.1007/s11214-008-9479-0}

\bibitem[{Bzowski {et~al.}(2013{\natexlab{a}})Bzowski, Sok{\'o}{\l}, Kubiak, \&
  Kucharek}]{bzowski_etal:13b}
Bzowski, M., Sok{\'o}{\l}, J.~M., Kubiak, M.~A., \& Kucharek, H.
  2013{\natexlab{a}}, \aap, 557, A50, \dodoi{10.1051/0004-6361/201321700}

\bibitem[{Bzowski {et~al.}(2013{\natexlab{b}})Bzowski, Sok{\'{o}}{\l},
  Tokumaru, Fujiki, Qu{\'e}merais, Lallement, Ferron, Bochsler, \&
  McComas}]{bzowski_etal:13a}
Bzowski, M., Sok{\'{o}}{\l}, J.~M., Tokumaru, M., {et~al.} 2013{\natexlab{b}},
  in {Cross-Calibration of Far {UV} Spectra of Solar Objects and the
  Heliosphere}, ed. E.~Qu{\'e}merais, M.~Snow, \& R.~Bonnet, {ISSI Scientific
  Report} No.~13 ({Springer Science+Business Media}), 67--138, doi
  10.1007/978--1--4614--6384--9$\_$3

\bibitem[{Fahr(1978)}]{fahr:78}
Fahr, H.~J. 1978, \aap, 66, 103

\bibitem[{Fahr(1979)}]{fahr:79}
---. 1979, \aap, 77, 101

\bibitem[{{Fuselier} {et~al.}(2009){Fuselier}, {Bochsler}, {Chornay}, {Clark},
  {Crew}, {Dunn}, {Ellis}, {Friedmann}, {Funsten}, {Ghielmetti}, {Googins},
  {Granoff}, {Hamilton}, {Hanley}, {Heirtzler}, {Hertzberg}, {Isaac}, {King},
  {Knauss}, {Kucharek}, {Kudirka}, {Livi}, {Lobell}, {Longworth}, {Mashburn},
  {McComas}, {M{\"o}bius}, {Moore}, {Moore}, {Nemanich}, {Nolin}, {O'Neal},
  {Piazza}, {Peterson}, {Pope}, {Rosmarynowski}, {Saul}, {Scherrer}, {Scheer},
  {Schlemm}, {Schwadron}, {Tillier}, {Turco}, {Tyler}, {Vosbury}, {Wieser},
  {Wurz}, \& {Zaffke}}]{fuselier_etal:09b}
{Fuselier}, S.~A., {Bochsler}, P., {Chornay}, D., {et~al.} 2009, \ssr, 146,
  117, \dodoi{10.1007/s11214-009-9495-8}

\bibitem[{{Galli} {et~al.}(2019){Galli}, {Wurz}, {Rahmanifard}, {M{\"o}bius},
  {Schwadron}, {Kucharek}, {Heirtzler}, {Fairchild}, {Bzowski}, {Kubiak},
  {Kowalska-Leszczy{\'n}ska}, {Sok{\'o}{\l}}, {Fuselier}, {Swaczyna}, \&
  {McComas}}]{galli_etal:19a}
{Galli}, A., {Wurz}, P., {Rahmanifard}, F., {et~al.} 2019, \apj, 871, 52,
  \dodoi{10.3847/1538-4357/aaf737}

\bibitem[{Katushkina {et~al.}(2015)Katushkina, Izmodenov, Alexashov, Schwadron,
  \& McComas}]{katushkina_etal:15b}
Katushkina, O.~A., Izmodenov, V.~V., Alexashov, D.~B., Schwadron, N.~A., \&
  McComas, D.~J. 2015, \apjs, 220, 33, \dodoi{10.1088/0067-0049-220-2-33}

\bibitem[{{Kowalska-Leszczynska} {et~al.}(2020){Kowalska-Leszczynska},
  {Bzowski}, {Kubiak}, \& {Sok{\'o}{\l}}}]{IKL:20a}
{Kowalska-Leszczynska}, I., {Bzowski}, M., {Kubiak}, M.~A., \& {Sok{\'o}{\l}},
  J.~M. 2020, \apjs, 247, 62, \dodoi{10.3847/1538-4365/ab7b77}

\bibitem[{{Kowalska-Leszczynska}
  {et~al.}(2018{\natexlab{a}}){Kowalska-Leszczynska}, {Bzowski},
  {Sok{\'o}{\l}}, \& {Kubiak}}]{IKL:18b}
{Kowalska-Leszczynska}, I., {Bzowski}, M., {Sok{\'o}{\l}}, J.~M., \& {Kubiak},
  M.~A. 2018{\natexlab{a}}, \apj, 868, 49, \dodoi{10.3847/1538-4357/aae70b}

\bibitem[{{Kowalska-Leszczynska}
  {et~al.}(2018{\natexlab{b}}){Kowalska-Leszczynska}, {Bzowski},
  {Sok{\'o}{\l}}, \& {Kubiak}}]{IKL:18a}
---. 2018{\natexlab{b}}, \apj, 852, 15, \dodoi{10.3847/1538-4357/aa9f2a}

\bibitem[{Kubiak {et~al.}(2021{\natexlab{a}})Kubiak, Bzowski,
  Kowalska-Leszczynska, \& Strumik}]{kubiak_etal:21a}
Kubiak, M.~A., Bzowski, M., Kowalska-Leszczynska, I., \& Strumik, M.
  2021{\natexlab{a}}, \apjs, 254, 16, \dodoi{10.3847/1538-4365/abeb79}

\bibitem[{Kubiak {et~al.}(2021{\natexlab{b}})Kubiak, Bzowski,
  Kowalska-Leszczynska, \& Strumik}]{kubiak_etal:21b}
---. 2021{\natexlab{b}}, \apjs, 254, 17, \dodoi{10.3847/1538-4365/abeb78}

\bibitem[{Kubiak {et~al.}(2016)Kubiak, Swaczyna, Bzowski, Sok{\'o}{\l},
  Fuselier, Galli, Heirtzler, Kucharek, Leonard, McComas, Park, Schwadron, \&
  Wurz}]{kubiak_etal:16a}
Kubiak, M.~A., Swaczyna, P., Bzowski, M., {et~al.} 2016, \apjs, 223, 35,
  \dodoi{10.1088/0067-0049/220/2/35}

\bibitem[{Lemaire {et~al.}(2015)Lemaire, Vial, Curdt, Sch{\"u}hle, \&
  Wilhelm}]{lemaire_etal:15a}
Lemaire, P., Vial, J., Curdt, W., Sch{\"u}hle, U., \& Wilhelm, K. 2015, \aap,
  581, A26

\bibitem[{Machol {et~al.}(2019)Machol, Snow, Woodraska, Woods, Viereck, \&
  Coddington}]{machol_etal:19a}
Machol, J.~L., Snow, M., Woodraska, D., {et~al.} 2019, Earth and Space Science,
  6, 2263, \dodoi{10.1029/2019EA000648}

\bibitem[{{McComas} {et~al.}(2009){McComas}, {Allegrini}, {Bochsler},
  {Bzowski}, {Collier}, {Fahr}, {Fichtner}, {Frisch}, {Funsten}, {Fuselier},
  {Gloeckler}, {Gruntman}, {Izmodenov}, {Knappenberger}, {Lee}, {Livi},
  {Mitchell}, {M{\"o}bius}, {Moore}, {Pope}, {Reisenfeld}, {Roelof},
  {Scherrer}, {Schwadron}, {Tyler}, {Wieser}, {Witte}, {Wurz}, \&
  {Zank}}]{mccomas_etal:09a}
{McComas}, D.~J., {Allegrini}, F., {Bochsler}, P., {et~al.} 2009, \ssr, 146,
  11, \dodoi{10.1007/s11214-009-9499-4}

\bibitem[{{McComas} {et~al.}(2011){McComas}, {Carrico}, {Hautamaki},
  {Intelisano}, {Lebois}, {Loucks}, {Policastri}, {Reno}, {Scherrer},
  {Schwadron}, {Tapley}, \& {Tyler}}]{mccomas_etal:11a}
{McComas}, D.~J., {Carrico}, J.~P., {Hautamaki}, B., {et~al.} 2011, Space
  Weather, 9, S11002, \dodoi{10.1029/2011SW000704}

\bibitem[{{Meier}(1977)}]{meier:77a}
{Meier}, R.~R. 1977, \aap, 55, 211

\bibitem[{Qu{\'e}merais(2000)}]{quemerais:00}
Qu{\'e}merais, E. 2000, \aap, 358, 353

\bibitem[{{Qu{\'e}merais}(2006)}]{quemerais:06a}
{Qu{\'e}merais}, E. 2006, in The Physics of the Heliospheric Boundaries, ed.
  {V.~V.~Izmodenov \& R.~Kallenbach}, 283--310

\bibitem[{Rahmanifard {et~al.}(2019)Rahmanifard, M{\"o}bius, Schwadron, Galli,
  Richards, Kucharek, Sok{\'{o}}{\l}, Heirtzler, Lee, Bzowski,
  Kowalska-Leszczynska, Kubiak, Wurz, Fuselier, \&
  McComas}]{rahmanifard_etal:19a}
Rahmanifard, F., M{\"o}bius, E., Schwadron, N.~A., {et~al.} 2019, \apj, 887,
  217, \dodoi{10.3847/1538-4357/ab58ce}

\bibitem[{Ruci{\'n}ski \& Bzowski(1995)}]{rucinski_bzowski:95b}
Ruci{\'n}ski, D., \& Bzowski, M. 1995, \aap, 296, 248

\bibitem[{{Scherer} \& {Fahr}(1996)}]{scherer_fahr:96}
{Scherer}, H., \& {Fahr}, H.~J. 1996, \aap, 309, 957

\bibitem[{{Schwadron} {et~al.}(2013){Schwadron}, {Moebius}, {Kucharek}, {Lee},
  {French}, {Saul}, {Wurz}, {Bzowski}, {Fuselier}, {Livadiotis}, {McComas},
  {Frisch}, {Gruntman}, \& {Mueller}}]{schwadron_etal:13a}
{Schwadron}, N.~A., {Moebius}, E., {Kucharek}, H., {et~al.} 2013, \apj, 775,
  86, \dodoi{10.1088/0004-637X/775/2/86}

\bibitem[{{Sok\'{o}{\l}} {et~al.}(2015){Sok\'{o}{\l}}, {Kubiak}, {Bzowski}, \&
  {Swaczyna}}]{sokol_etal:15b}
{Sok\'{o}{\l}}, J.~M., {Kubiak}, M.~A., {Bzowski}, M., \& {Swaczyna}, P. 2015,
  \apjs, 220, 27, \dodoi{10.1088/0067-0049/220/2/27}

\bibitem[{{Sok{\'o}{\l}} {et~al.}(2020){Sok{\'o}{\l}}, {McComas}, {Bzowski}, \&
  {Tokumaru}}]{sokol_etal:20a}
{Sok{\'o}{\l}}, J.~M., {McComas}, D.~J., {Bzowski}, M., \& {Tokumaru}, M. 2020,
  \apj, 897, 179, \dodoi{10.3847/1538-4357/ab99a4}

\bibitem[{{Strumik} {et~al.}(2021){Strumik}, {Bzowski}, \&
  {Kubiak}}]{strumik_etal:21b}
{Strumik}, M., {Bzowski}, M., \& {Kubiak}, M.~A. 2021, \apjl, 919, L18,
  \dodoi{10.3847/2041-8213/ac2734}

\bibitem[{Swaczyna {et~al.}(2020)Swaczyna, McComas, Zirnstein, Sok{\'o}{\l},
  Elliott, Bzowski, Kubiak, Richardson, Kowalska-Leszczynska, Stern, Weaver,
  Olkin, Singer, \& Spencer}]{swaczyna_etal:20a}
Swaczyna, P., McComas, D., Zirnstein, E.~J., {et~al.} 2020, \apj, 903, 48,
  \dodoi{10.3847/1538-4357/abb80a}

\bibitem[{Tarnopolski \& Bzowski(2009)}]{tarnopolski_bzowski:09}
Tarnopolski, S., \& Bzowski, M. 2009, \aap, 493, 207,
  \dodoi{10.1051/0004-6361:20077058}

\bibitem[{Thomas(1978)}]{thomas:78}
Thomas, G.~E. 1978, Ann. Rev. Earth Planet. Sci., 6, 173

\bibitem[{{Wu} \& {Judge}(1979)}]{wu_judge:79b}
{Wu}, F.~M., \& {Judge}, D.~L. 1979, \jgr, 84, 979,
  \dodoi{10.1029/JA084iA03p00979}

\bibitem[{Wu \& Judge(1979)}]{wu_judge:79a}
Wu, F.~M., \& Judge, D.~L. 1979, \apj, 231, 594

\end{thebibliography}

\end{document}